\pdfoutput=1

\documentclass[11pt]{article}

\usepackage{acl}

\usepackage{times}
\usepackage{latexsym}

\usepackage[T1]{fontenc}

\usepackage[utf8]{inputenc}

\usepackage{microtype}

\usepackage{inconsolata}

\usepackage{adjustbox}
\usepackage{multirow}
\usepackage{multicol}
\usepackage{booktabs}
\usepackage{setspace}
\usepackage{hyperref}
\usepackage{threeparttable}
\usepackage{xcolor}
\usepackage{colortbl}
\usepackage{float}
\usepackage{enumitem}
\usepackage{array}
\usepackage{arydshln}
\usepackage{makecell}
\newcommand{\method}{{BioT5+}}

\newcommand{\text}{$\langle$\texttt{Text Description}$\rangle$}
\newcolumntype{P}[1]{>{\centering\arraybackslash}p{#1}}

\title{BioT5+: Towards Generalized Biological Understanding with IUPAC Integration and Multi-task Tuning}

\author{
    Qizhi Pei\textsuperscript{1},
    {\bf Lijun Wu\textsuperscript{2}$^{\ast}$},
    Kaiyuan Gao\textsuperscript{3},
    Xiaozhuan Liang\textsuperscript{4},
    Yin Fang\textsuperscript{4},\\
    {\bf Jinhua Zhu\textsuperscript{5}},
    {\bf Shufang Xie\textsuperscript{1}},
    {\bf Tao Qin\textsuperscript{2}},
    {\bf Rui Yan\textsuperscript{1,6}\thanks{\ \ Corresponding authors: Lijun Wu (\url{apeterswu@gmail.com}) and Rui Yan (\url{ruiyan@ruc.edu.cn})}} \\
    \textsuperscript{1}Gaoling School of Artificial Intelligence, Renmin University of China \\
    \textsuperscript{2}Microsoft Research \quad
    \textsuperscript{3}Huazhong University of Science and Technology\\
    \textsuperscript{4}Zhejiang University \quad
    \textsuperscript{5}University of Science and Technology of China \\
    \textsuperscript{6}Engineering Research Center of Next-Generation Intelligent Search\\ and Recommendation, Ministry of Education \\
    \texttt{\{qizhipei,shufangxie,ruiyan\}@ruc.edu.cn} \quad
    \texttt{apeterswu@gmail.com} \\
    \texttt{im\_kai@hust.edu.cn} \quad
    \texttt{\{liangxiaozhuan,fangyin\}@zju.edu.cn} \\
    \texttt{teslazhu@mail.ustc.edu.cn} \quad
    \texttt{taoqin@microsoft.com} \\
}

\begin{document}
\maketitle
\begin{abstract}
Recent research trends in computational biology have increasingly focused on integrating text and bio-entity modeling, especially in the context of molecules and proteins. However, previous efforts like BioT5 faced challenges in generalizing across diverse tasks and lacked a nuanced understanding of molecular structures, particularly in their textual representations (e.g., IUPAC). This paper introduces BioT5+, an extension of the BioT5 framework, tailored to enhance biological research and drug discovery. BioT5+ incorporates several novel features: integration of IUPAC names for molecular understanding, inclusion of extensive bio-text and molecule data from sources like bioRxiv and PubChem, the multi-task instruction tuning for generality across tasks, and a numerical tokenization technique for improved processing of numerical data. These enhancements allow BioT5+ to bridge the gap between molecular representations and their textual descriptions, providing a more holistic understanding of biological entities, and largely improving the grounded reasoning of bio-text and bio-sequences. The model is pre-trained and fine-tuned with a large number of experiments, including \emph{3 types of problems (classification, regression, generation), 15 kinds of tasks, and 21 total benchmark datasets}, demonstrating the remarkable performance and state-of-the-art results in most cases. BioT5+ stands out for its ability to capture intricate relationships in biological data, thereby contributing significantly to bioinformatics and computational biology. Our code is available at \url{https://github.com/QizhiPei/BioT5}.
\end{abstract}

\section{Introduction}

Molecules and proteins are two crucial bio-entities in drug discovery, forming the foundation of biological activities~\citep{ml_drug_discovery,ai4science2023impact}. 
A molecule can be represented by its SMILES~\citep{weininger1988smiles,weininger1989smiles} or SELFIES~\citep{krenn2020self} sequence, and a protein can be described by a FASTA~\citep{lipman1985rapid,pearson1988improved} sequence. 
With the advancement of Language Models (LMs), an increasing body of work focuses on understanding the molecules and proteins by modeling their bio-sequences~\citep{chemberta,esm,esm2}.

Notably, biological literature~\citep{pubmed1.0,pubmed2.0} is full of extensive information on molecules and proteins. 
When a biological entity is mentioned in such literature, its context is predominantly centered around a description of some characteristics of the entity. 
Consequently, there has been a growing body of work dedicated to the joint modeling of text and biological entities~\cite{pei2024leveraging}, such as Galactica~\citep{galactica}, MolXPT~\citep{molxpt}, BioT5~\citep{biot5} and BioMedGPT~\citep{biomedgpt}, which are all scientific models trained on text, molecule and protein sequences. 
Despite their achievements, substantial opportunities for enhancement still remain:
(1) Prior works neglect the importance of modeling the textual name of molecules, such as International Union of Pure and Applied Chemistry (IUPAC), which provides a standard and systematic naming method for ensuring uniformity and clarity across the scientific community. 
Different from SMILES and SELFIES, IUPAC bears a closer resemblance to natural language that is evident in its widespread adoption within scientific literature~\citep{iupac_detect}.
(2) Previous models were predominantly specialist models, necessitating the training of a separate model for each downstream task, thereby lacking in generality and increasing the training and developing cost~\cite{molxpt,biot5}.
(3) Most of the previous models based on T5~\citep{t5} and GPT~\citep{gpt3} architectures only focus on the classification tasks since they do not implement specialized tokenization for numerical data, which results in their suboptimal adaptation to regression tasks.

To address the above challenges, in this paper, we introduce \method, an advanced iteration of the BioT5 framework~\citep{biot5}, designed to augment biological research and drug discovery with enriched data integration, multi-task capabilities, and the ability to solve regression tasks. Shortly speaking, \method{} incorporates following significant enhancements:

(1) {\em Enhanced Molecule Understanding}: By integrating IUPAC name into \method{} framework, the model can achieve a deeper comprehension of molecular structures. This integration allows \method{} to interpret chemical names as they commonly appear in scientific literature, bridging the gap between formal molecular representations (such as SELFIES) and their textual descriptions. Consequently, this enhances the understanding of molecules and facilitates more accurate predictions and analyses of molecular properties and activities.

(2) {\em Expanded Bio-text and Molecule Data}: Compared to BioT5, \method{} includes an extensive corpus of bio-text data from sources like bioRxiv~\cite{sever2019biorxiv} and PubMed~\cite{pubmed1.0,pubmed2.0}, alongside high-quality molecular data from PubChem~\cite{kim2019pubchem}. This expansion not only broadens the knowledge base of the model but also enriches the contextual understanding of biological entities.

(3) {\em Multi-task Instruction Tuning}: \method{} employs multi-task instruction tuning strategy for downstream tasks rather than the separate specialized model training for each task. By leveraging a unified and multi-task training framework, \method{} can seamlessly integrate knowledge from diverse tasks, enhancing its predictive power and generalization capabilities across different biological and chemical domains.

(4) {\em Advanced Numerical Tokenization}: To overcome the limitations of the numerical representations, \method{} integrates an advanced character-based numerical tokenization strategy, drawing inspiration from the Llama~\citep{llama} model.  
This technique allows for a more nuanced and consistent representation of numerical values.

With our designed pre-training and multi-task instruction tuning, the effectiveness of \method{} is verified on \emph{3 types of problems (classification, generation, and regression), 15 different tasks, and 21 benchmark datasets}, including molecule property prediction, retrosynthesis, molecule description generation, drug-target interaction, and so on.
\method{} has shown highly competitive results, achieving state-of-the-art performance in most of the tasks. 
This robust performance underscores the enhanced capability of \method{} to capture and analyze the intricate relationships and properties inherent in biological data, marking a significant step forward in computational biology.

\begin{figure*}
    \centering
    \vspace{-1.3cm}
    \includegraphics[width=1\linewidth]{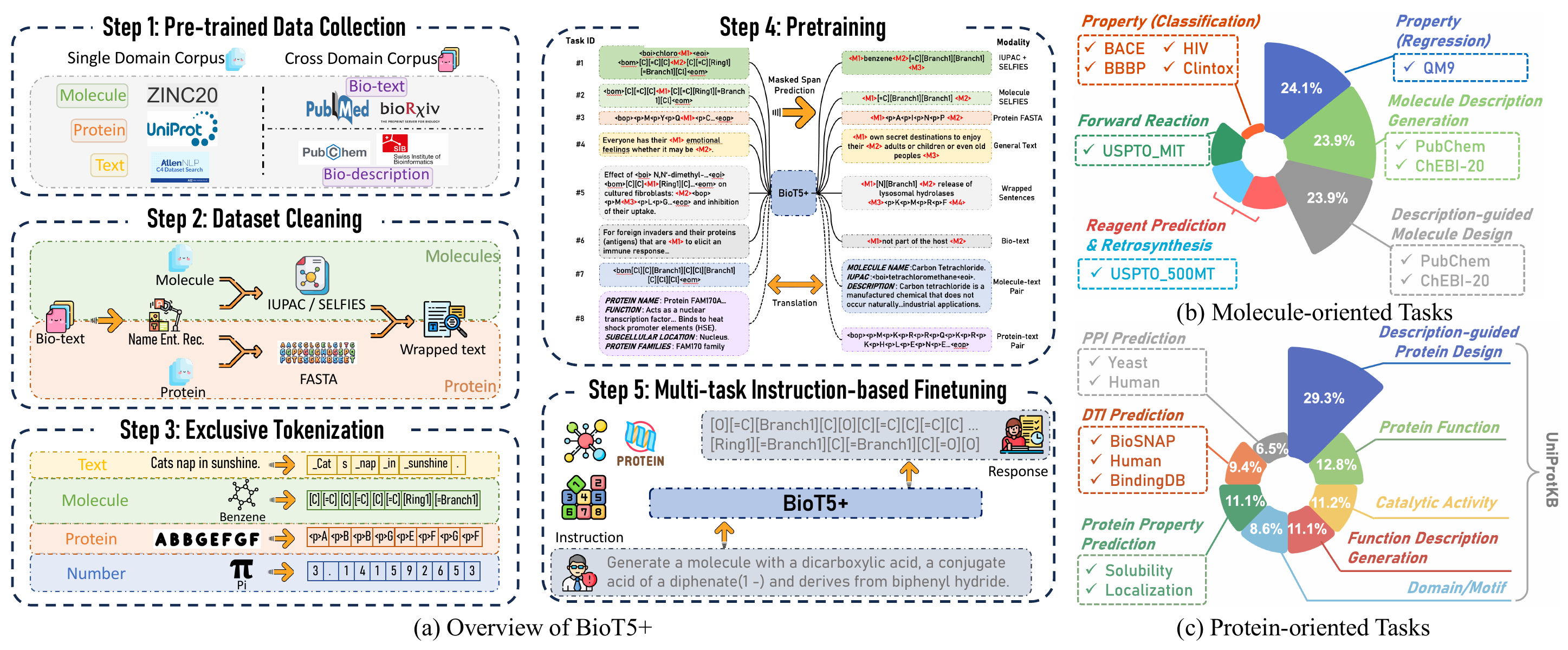}
    \vspace{-0.8cm}
    \caption{\footnotesize{
    (a): The overview of \method{} framework. 
    (b) (c): the composition of \method{} downstream tasks, which is divided into two categories: 
    (b) molecule-oriented tasks and (c) protein-oriented tasks. The names of the tasks, along with their instruction datasets and respective percentages, are annotated near each segment of the accompanying pie charts.}}
    \label{fig:overview}
    \vspace{-0.5cm}
\end{figure*}
\section{Related Work}
\subsection{Biological Cross-modal Models}
Recent advancements in LLMs have led to an increased focus on jointly modeling molecules, proteins, and text, aiming to enhance the understanding of bio-entities through text. 

\noindent{\textbf{Molecule-Text.}}
MolT5~\citep{molt5} is jointly trained on general text and molecule SMILES using T5~\citep{t5} masked span prediction objective.
MoMu~\citep{momu} employs contrastive learning on molecular graphs and related text, and MolFM~\citep{molfm} further incorporates knowledge graph embedding for molecule representation.
MolXPT~\citep{molxpt} is jointly trained on molecule SMILES and wrapped text using GPT~\cite{gpt3} framework.
MolCA~\citep{molca} enhances LMs by integrating 2D molecular graph perception through a cross-modal projector and uni-modal adapter.
GIT-Mol~\citep{git-mol} is a multi-modal LLM that synergizes graphs, images, SMILES, and molecule captions.
Text+Chem T5~\cite{text+chemt5} is a multi-domain, multi-task language model capable of concurrently processing molecules and natural language.

\noindent{\textbf{Protein-Text.}}
Several notable works focus on jointly modeling proteins and text.
ProteinDT~\citep{proteindt} presents a text-guided protein design framework. 
BioTranslator~\citep{biotranslator} is a cross-modal translation system, which can annotate various biological instances using textual descriptions.
Prot2Text~\citep{prottext} combines GNN and LLM in an encoder-decoder framework to generate protein functions in a free-text style.

In addition to the models mentioned above, there are other models trained in a more diverse range of modalities:
DeepEIK~\citep{luo2023empowering} is a multi-modal model which integrates features from multi-modal inputs including drugs, proteins, and text.
BioT5~\citep{biot5} is a T5-based~\citep{t5} model that undergoes joint training on text, molecule SELFIES, and protein FASTA sequences, effectively bridging the gap between textual and biological data.

Despite the successes of these models, their focus on single-task training limits their versatility and hinders the development of a more generalized and adaptable approach in computational biology.

\subsection{Instruction Tuning for Biological Tasks}
Instruction tuning is a popular technique applied to pre-trained LLMs where they are trained with specialized instruction datasets, thus equipping LLMs with the ability to understand task-specific instructions.
Recently, there has been a growing interest in exploring instruction tuning for various biological tasks. Notable among these efforts is the development of Mol-Instructions~\cite{mol-instructions}, a comprehensive instruction dataset specifically designed for the biological domain, which includes molecule-oriented instructions, protein-oriented instructions, and biomolecular text instructions.
InstructMol~\cite{instructmol}, a multi-modal LLM, employs instruction tuning to align molecular graphs, molecule SELFIES, and natural language.
Different from these approaches, our \method{} is specifically pre-trained for the biological domain. Further instruction tuning within biological instructions enables \method{} to not only understand bio-entities but also generalize across various biological tasks.

\section{BioT5+ Framework}
This section introduces \method{} framework and an overview is shown in Figure~\ref{fig:overview}.
The tasks involved in pre-training are presented in Figure~\ref{fig:pipeline}.

\subsection{Intuition for IUPAC Integration}
IUPAC naming system provides a standardized set of rules for naming chemical compounds, which allows for the precise description of molecular structures and their components (functional groups, chains, and rings), making it a cornerstone in chemical nomenclature.
Typically, an IUPAC name is constructed from the names of individual molecules' constituent parts, reflecting their structure. This includes prefixes, infixes, and suffixes that indicate various chemical groups and structural features, providing a comprehensive description of the molecule. 
For instance, the IUPAC name for Aspirin is ``2-acetyloxybenzoic acid''. Here, ``2-acetyloxy'' refers to an acetoxy group attached to the second carbon of a benzene ring, and ``benzoic acid'' indicates a benzene ring with a carboxylic acid group.
The resemblance of IUPAC names to natural language, coupled with their prevalent use in scientific literature, makes them an ideal candidate for model pre-training. By pre-training the model on literature that includes IUPAC names, \method{} can establish a nuanced understanding of the relationship between molecules and various textual descriptions of their chemical properties.

\subsection{Pre-training Corpus}
\label{pretrain_corpus}
As an extension of BioT5, the majority of the pre-training corpus for \method{} is identical to BioT5, hence we will briefly mention the common elements while focusing primarily on the novel aspects introduced in \method.

The pre-training corpus consists of 4 classes:
(1) {\em Single-modal data}, including molecule SELFIES with IUPAC name from PubChem~\citep{kim2019pubchem}, molecule SELFIES from ZINC20~\citep{irwin2020zinc20}, protein FASTA from Uniref50~\citep{suzek2007uniref}, and general text from ``Colossal Clean Crawled Corpus'' (C4)~\citep{t5}. 
For the molecule from PubChem, we concatenate the IUPAC name and SELFIES for pre-training as shown in Figure~\ref{fig:pipeline}.
(2) {\em Wrapped text}, where the molecule or gene/protein names are suffixed with corresponding sequence representation. 
We employ BERN2~\citep{sung2022bern2}, a neural-based Named Entity Recognition (NER) system in the biological domain, to detect and classify occurrences of molecules and proteins within the abstracts of PubMed~\citep{pubmed2.0} and bioRxiv~\citep{sever2019biorxiv}.
For the molecule name, we first standardize the name to its IUPAC name and then append the corresponding SELFIES. For the gene/protein name, we will directly append its FASTA sequence. 
For the generation of high-quality wrapped text, we also analyze the confidence score distribution predicted by BERN2~\citep{sung2022bern2}. Only those entities with higher confidence scores were retained to ensure the accuracy and relevance of the appended sequence data. Further detailed descriptions of this process are provided in Appendix Section~\ref{appsec:ner_entity_link}.
(3) {\em Bio-text}, including PubMed~\citep{pubmed2.0} Central full text articles, and bio-texts from PubMed~\citep{pubmed2.0} abstracts and bioRxiv~\citep{sever2019biorxiv} abstracts that do not yield identifiable named entities in (2).
(4) {\em Molecule-description pairs} and {\em protein-description pairs}. The molecule-text data is collected from PubChem~\citep{kim2019pubchem} and we also add the IUPAC name to the text description. All molecules and proteins that exist in the downstream Mol-Instructions dataset~\citep{mol-instructions} and ChEBI-20~\citep{molt5} are excluded to prevent data leakage. The protein-text data is the same as BioT5~\citep{biot5}.

\noindent{\textbf{Remarkable Difference.}} The primary distinctions between \method{} and BioT5 are as follows:
(1) \method{} integrates IUPAC in the molecular pre-training data, encompassing IUPAC names combined with SELFIES, wrapped text, and molecule-text translation data. More details are in Appendix Section~\ref{appsec:iupac}.
(2) \method{} incorporates a broader spectrum of high-quality data, including IUPAC names and SELFIES from PubChem, as well as comprehensive articles from bioRxiv and PubMed Central.

\subsection{Tokenization}
BioT5~\citep{biot5} has already demonstrated the advantages of employing separate tokenization and embedding techniques. 
\method{} inherits this advantage to apply specialized tokenization and vocabulary specifically for bio-entities. This method explicitly differentiates between the biological semantic space and the textual semantic space. 
For molecule SELFIES, each chemically meaningful atom group, naturally distinguished from textual vocabulary due to its bracketed format like [C], is tokenized as an individual token using the inherent token set defined by SELFIES. 
For protein FASTA sequences, to ensure a clear modal distinction, each amino acid is tokenized into a separate token with the prefix <p>, differentiating them from standard upper-case English letters.

Concurrently, the tokenization of numerical data is worth dedicated consideration and design. The direct application of T5~\citep{t5} dictionary derived from nature language using SentencePiece~\citep{DBLP:conf/emnlp/KudoR18} for numerical tokenization can lead to inconsistencies~\citep{goat}. 
For instance, the number $1024$ might be tokenized into ``$10$'' and ``$24$'', while ``$2048$'' could be split into ``$2$'', ``$0$'', and ``$48$''.
This irregular segmentation poses a challenge for the model in consistently mapping embeddings to numbers, especially when the number of digits they represent varies.
In contrast, models like Llama~\citep{llama,llama2} and ChatGLM~\citep{chatglm} adopt a character-based approach to numerical tokenization, where each digit is tokenized as an individual token. This method has been demonstrated to yield superior results in various arithmetic tasks~\citep{goat,DBLP:journals/corr/abs-2102-13019}. Accordingly, in \method{} we also implement this character-based approach for numerical tokenization without modifying the original dictionary. 
The efficacy of this method over the original T5~\citep{t5} and BioT5~\citep{biot5} to numerical tokenization are shown in Section~\ref{sec:ablation}, providing empirical evidence of its superior performance in handling numerical data.

\subsection{Model and Training}
\noindent{\bf Model architecture.} \method{} adopts the same architecture as the BioT5~\citep{biot5}, which follows the T5-v1.1-base\footnote{\url{https://huggingface.co/docs/transformers/model_doc/t5v1.1}} configuration with vocab size $35,076$ and $252$M parameters.

\noindent{\bf Pre-training.} Based on the pre-training corpus described in Section~\ref{pretrain_corpus}, the pre-training for \method{} is conducted in a multi-task way with eight tasks that fall into 4 categories:
(1) {\em Modality-Specific T5 Objectives}: This category involves the application of the T5 objective (masked span prediction) to each modality in isolation, including molecule SELFIES with IUPAC name (Task \#1), molecule SELFIES (Task \#2), protein FASTA sequences (Task \#3), and general textual content (Task \#4). 
(2) {\em T5 Objectives on Wrapped Text}: Applying the T5 objective to ``wrapped'' text extracted from scientific corpora (Task \#5).
(3) {\em T5 Objectives on Bio-text}: Applying the T5 objective to text in biological domain (Task \#6).
(4) {\em Bidirectional Translation Tasks}: This involves the bidirectional translation between molecule SELFIES-text pairs (Task \#7) and protein FASTA-text pairs (Task \#8). 
Through these strategically structured pre-training tasks, \method{} is adept at learning the intricate relationships and characteristics of bio-entities as represented in textual information. 

\noindent{\bf Multi-task Instruction-based Fine-tuning.} After the comprehensive pre-training phase, \method{} undergoes multi-task instruction-based fine-tuning.
Unlike BioT5 where each downstream task has a specialized fine-tuned model, we follow~\citealp{mol-instructions} and ~\citealp{instructmol} to categorize downstream tasks and conduct multi-task instruction tuning, which not only saves the repeated tuning cost but also eases the model deployment for the evaluation of multiple tasks. 
The relevant groupings and information about benchmark tasks and datasets are illustrated in Figure~\ref{fig:overview}, which is simply split by the domains, e.g., molecule-oriented or protein-oriented tasks.
This methodology serves a dual purpose: firstly, it bridges the gap between the pre-training and fine-tuning phases, ensuring a smoother transition and integration of learned capabilities. Secondly, it activates and harnesses the general capabilities of \method{} across various tasks, demonstrating its versatility and adaptability in handling diverse biological problems.

\section{Experiments and Results}
As shown in Figure~\ref{fig:overview}, \method{} is extensively evaluated across 21 well-established downstream benchmark datasets, which can be classified into 7 molecule-oriented tasks and 8 protein-oriented tasks with 3 types of problems: classification, regression, and generation. 
Following~\citealp{mol-instructions}, we categorize downstream tasks into different groups for multi-task instruction tuning in the same way, and details about the downstream datasets and baselines are in Appendix Section~\ref{appsec:finetune_detail}.

\subsection{Molecule-oriented Tasks}
The molecule-oriented tasks cover different topics. 
As we incorporate IUPAC name for molecule in the pre-training, we also use IUPAC in some molecule-oriented tasks, such as molecule property prediction and molecule description generation. More details are in the following sections and Appendix.
\begin{table}
\centering
\caption{\footnotesize{
Performance (AUROC) comparison on molecule property prediction tasks (classification) on MoleculeNet~\cite{moleculenet} benchmark (\textbf{Best}, \underline{Second Best}).
$^*$ means LoRA~\cite{lora} tuning.}
}
\small
\resizebox{\linewidth}{!}{
\begin{tabular}{lcccc}
    \toprule
    \textsc{Method} &BACE $\uparrow$  &BBBP $\uparrow$ &HIV $\uparrow$ &Clintox $\uparrow$\\
    \textsc{\# Molecules} &1513 &2039 &41127 &1478\\
    \midrule
    \rowcolor[RGB]{234, 238, 234} \multicolumn{5}{l}{\textit{Single-task Specialist Models}} \\
    GraphCL &75.4 &69.7 &78.5 &76.0\\ 
    GraphMVP-C &81.2 &72.4 &77.0 &77.5 \\ 
    MGSSL & 79.7 & 70.5 & 79.5 & 80.7 \\ 
    MolCLR & \underline{89.0} & 73.8 & 80.6 & 93.2 \\ 
    GEM & 85.6 & 72.4 & 80.6 & 90.1 \\ 
    Uni-Mol & 85.7 & 72.9 & \underline{80.8} & 91.9\\ 
    KV-PLM &71.9 &66.9 &68.8 &84.3\\ 
    MoMu &76.7 &70.5 & 75.9 &79.9\\ 
    MolFM &83.9 & 72.9 & 78.8 & 79.7 \\ 
    MolXPT & 88.4 & \textbf{80.0} & 78.1 & \underline{95.3} \\ 
    BioT5 & \textbf{89.4} & \underline{77.7} & \textbf{81.0} & \textbf{95.4} \\ 
    \midrule 
    
    \rowcolor[RGB]{234, 238, 234} \multicolumn{5}{l}{\textit{LLM-based Generalist Models}} \\
    Galactica-6.7B & 58.4 & 53.5 & 72.2 & 78.4\\ 
    Galactica-30B  & 72.7 & 59.6 & 75.9 & 82.2 \\ 
    Galactica-120B & 61.7 & 66.1 & 74.5 & \underline{82.6}\\ 
    Vicuna-v1.5-13B-16k (4-shot) & 49.2 & 52.7 & 50.5 & -\\
    Vicuna-v1.3-7B$^*$  & 68.3 & 60.1 & 58.1 & -\\
    Llama-2-7B-chat$^*$ & 74.8 & 65.6 & 62.3 & -\\
    InstructMol-G-6.9B & \underline{85.9} & 64.0 & \underline{74.0} & -\\
    InstructMol-GS-6.9B & 82.3 & \underline{70.0} & 68.9 & -\\
    \midrule
    \textbf{\method} & \textbf{86.2} & \textbf{76.5} & \textbf{76.3} & \textbf{92.3} \\
    \bottomrule
\end{tabular}
}
\vspace{-0.7cm}
\label{tab:mol_property_pred_classification}
\end{table}

\begin{table*}[t]
\centering
\vspace{-1.3cm}
\small
\caption{\footnotesize{Performance comparison on chemical reaction-related tasks (\textbf{Best}, \underline{Second Best}). 
$*$ means LoRA tuning.}
}
\setlength{\tabcolsep}{4.3mm}{
\scalebox{0.65}{
\begin{tabular}{lccccccc}
\toprule
\textsc{Model}
&\textsc{Exact}$\uparrow$  & \textsc{BLEU}$\uparrow$  & \textsc{Levenshtein}$\downarrow$  & \textsc{RDK FTS}$\uparrow$  & \textsc{MACCS FTS}$\uparrow$ & \textsc{Morgan FTS}$\uparrow$ & \textsc{Validity}$\uparrow$ \\
\midrule[1.1pt]
\rowcolor[RGB]{234, 238, 234}
\multicolumn{8}{l}{\textit{Reagent Prediction}} \\
Llama-7B & 0.000 & 0.003 & 28.040 & 0.037 & 0.001 & 0.001 & 0.001 \\
Galactica-6.7B & 0.000 & 0.141 & 30.760 & 0.036 & 0.127 & 0.051 & 0.995 \\
Text+Chem T5-223M & 0.000 & 0.225 & 49.323 & 0.039 & 0.186 & 0.052 & 0.313 \\
Mol-Instructions-7B & 0.044 & 0.224 & 23.167 & 0.237 & 0.364 & 0.213 & 1.000 \\
Llama-7B$^*$(LoRA) &0.000	&0.283	&53.510	&0.136	&0.294	&0.106	&1.000 \\
InstructMol-G-6.9B & 0.070 & \textbf{0.890} &24.732 & \underline{0.469} & \textbf{0.691}	& \underline{0.426}	&1.000 \\
InstructMol-GS-6.9B & \underline{0.129} & 0.610 & \underline{19.664}& 0.444	&0.539	&0.400	&1.000 \\
\midrule
\textbf{\method} & \textbf{0.257} & \underline{0.695} & \textbf{12.901} & \textbf{0.539} & \underline{0.621} & \textbf{0.512} & 1.000 \\
\midrule[1.1pt]
\rowcolor[RGB]{234, 238, 234}
\multicolumn{8}{l}{\textit{Forward Reaction Prediction}} \\
Llama-7B & 0.000 & 0.020 & 42.002 & 0.001 & 0.002 & 0.001 & 0.039 \\
Galactica-6.7B & 0.000 & 0.468 & 35.021 & 0.156 & 0.257 & 0.097 & 0.946 \\
Text+Chem T5-223M & 0.239 & 0.782 & 20.413 & 0.705 & 0.789 & 0.652 & 0.762 \\
Mol-Instructions-7B & 0.045 & 0.654 & 27.262 & 0.313 & 0.509 & 0.262 & 1.000 \\
Llama-7B$^*$(LoRA) &0.012	&0.804	&29.947	&0.499	&0.649	&0.407	&1.000 \\
InstructMol-G-6.9B & 0.153 &0.906 &20.155	&0.519	&0.717	&0.457	&1.000 \\
InstructMol-GS-6.9B & \underline{0.536} & \underline{0.967} & \underline{10.851} & \underline{0.776} & \underline{0.878} & \underline{0.741}	&1.000\\
\midrule
\textbf{\method} & \textbf{0.864} & \textbf{0.993} & \textbf{3.403} & \textbf{0.949} & \textbf{0.975} & \textbf{0.935} & 1.000 \\
\midrule[1.1pt]
\rowcolor[RGB]{234, 238, 234}
\multicolumn{8}{l}{\textit{Retrosynthesis}} \\
Llama-7B & 0.000 & 0.036 & 46.844 & 0.018 & 0.029 & 0.017 & 0.010 \\
Galactica-6.7B & 0.000 & 0.452 & 34.940 & 0.167 & 0.274 & 0.134 & 0.986 \\
Text+Chem T5-223M & 0.141 & 0.765 & 24.043 & 0.685 & 0.765 & 0.585 & 0.698 \\
Mol-Instructions-7B & 0.009 & 0.705 & 31.227 & 0.283 & 0.487 & 0.230 & 1.000 \\
Llama-7B$^*$(LoRA) & 0.000 &0.283	&53.510	&0.136	&0.294	&0.106	&1.000 \\
InstructMol-G-6.9B & 0.114 & 0.586 & 21.271 & 0.422 & 0.523 &0.285	&1.000\\
InstructMol-GS-6.9B & \underline{0.407} & \underline{0.941} & \underline{13.967} & \underline{0.753} & \underline{0.852} & \underline{0.714} & 1.000\\
\midrule
\textbf{\method} & \textbf{0.642} & \textbf{0.969} & \textbf{6.710} & \textbf{0.897} & \textbf{0.930} & \textbf{0.866} & 1.000 \\
\bottomrule
\end{tabular}
}}
\vspace{-0.5cm}
\label{tab:chemical_reaction}
\end{table*}

\begin{table}
\centering
\small
\caption{\footnotesize{
Performance (MAE) comparison on molecule property prediction tasks (regression) on QM9 dataset from MoleculeNet~\cite{moleculenet} benchmark (\textbf{Best}, \underline{Second Best}).
$\Delta{\epsilon}$ means HOMO-LUMO gap.
}}
\scalebox{0.80}{
    \begin{tabular}{lccccc}
        \toprule
        \textsc{Method} &\textsc{HOMO} $\downarrow$ &\textsc{LUMO} $\downarrow$  &$\Delta{\epsilon}$ $\downarrow$ &\textsc{Avg} $\downarrow$\\
        \midrule 
        \rowcolor[RGB]{234, 238, 234} \multicolumn{5}{l}{\textit{LLM-based Generalist Models}} \\
        Llama2-7B (5-shot ICL) & 0.7367 & 0.8641 & 0.5152 & 0.7510 \\
        Vicuna-13B (5-shot ICL) & 0.7135 & 3.6807 & 1.5407 & 1.9783 \\
        Mol-Instructions-7B & 0.0210 & 0.0210 & 0.0203 & 0.0210 \\
        InstructMol-G-6.9B &  0.0060 & 0.0070 & 0.0082 & 0.0070 \\
        InstructMol-GS-6.9B & \underline{0.0048} & \underline{0.0050} & \underline{0.0061} & \underline{0.0050}\\
        \midrule
        \textbf{\method} & \textbf{0.0022} & \textbf{0.0024} & \textbf{0.0028} & \textbf{0.0025} \\
        \bottomrule
    \end{tabular}
}
\vspace{-0.5cm}
\label{tab:mol_property_pred_regression}
\end{table}

\subsubsection{Molecule Property Prediction}
Molecule property prediction is a crucial task in bioinformatics, focusing on the determination of specific properties exhibited by a given molecule. 
Following~\citealp{instructmol}, we explore the ability of \method{} on MoleculeNet~\citep{moleculenet} benchmark.
For classification tasks, we focus on 4 benchmark datasets: BACE, BBBP, HIV, and Clintox. 
Each sample includes an instruction detailing the property to be predicted and the molecule SELFIES with IUPAC name, with models required to generate a simple ``yes'' or ``no'' prediction.
For regression tasks, we focus on 3 regression benchmarks from QM9 dataset, which aims to predict quantum mechanical properties of molecules, based on the molecule SELFIES with IUPAC name, including HUMO, LUMO, and the HUMO-LUMO gap.

\noindent{\textbf{Results.}} The results for classification and regression tasks are shown in Table~\ref{tab:mol_property_pred_classification} and Table~\ref{tab:mol_property_pred_regression} respectively.
\method{} demonstrates superior performance over other generalist model baselines. 
Notably, for classification tasks, \method{} surpasses models like Galactica~\citep{galactica}, which is extensively trained on a vast corpus of scientific literature. 
Similarly, InstructMol~\citep{instructmol}, despite its inclusion of 2D graph information and LLMs, \method{} outperforms on both classification and regression tasks. 
This enhanced performance can be attributed to the integration of IUPAC names, wrapped text, bio-text, and molecule-text pairs in \method{} pre-training. The presence of molecule property descriptions in the context of these diverse corpora allows the model to acquire a comprehensive understanding of molecular properties.
However, when compared to single-task specialist models, \method{} showed some gaps. This discrepancy is understandable and can be attributed partly to the ease of tuning inherent in single-task models and partly to the fact that some baselines incorporated additional molecular information, such as 2D and 3D structures. 
\begin{table*}[t]
\centering
\vspace{-1.3cm}
\small
\caption{\footnotesize{Performance comparison on molecule description generation task on ChEBI-20~\cite{molt5} dataset.}
}
\setlength{\tabcolsep}{5.2mm}{
\scalebox{0.65}{
\begin{tabular}{lcccccc}
\toprule
\textsc{Model}
&\textsc{BLEU-2}$\uparrow$  & \textsc{BLEU-4}$\uparrow$  & \textsc{ROUGE-1}$\uparrow$  & \textsc{ROUGE-2}$\uparrow$  & \textsc{ROUGE-L}$\uparrow$ & \textsc{METEOR}$\uparrow$ \\
\midrule
\rowcolor[RGB]{234, 238, 234} \multicolumn{7}{l}{\textit{Single-task Specialist Models}} \\
Transformer & 0.061 & 0.027 & 0.204 & 0.087 & 0.186 & 0.114 \\
T5-base & 0.511 & 0.423 & 0.607 & 0.451 & 0.550 & 0.539 \\
MoT5-base & 0.540 &0.457 &0.634 &0.485 &0.568 &0.569 \\
MoMu (MolT5-base) & 0.549 &0.462 &  -    &  -    &  -    &0.576 \\
MolFM (MolT5-base) &0.585 &0.498 &0.653 &0.508 &0.594 &0.607 \\
MolXPT         &0.594  &0.505 &0.660 &0.511 &0.597 &0.626 \\
GIT-Mol-graph  & 0.290 &0.210 &0.540 &0.445 &0.512 & 0.491 \\
GIT-Mol-SMILES & 0.264 &0.176 &0.477 &0.374 &0.451 &0.430  \\
GIT-Mol-(graph+SMILES) & 0.352 & 0.263 &0.575 &0.485 &0.560 &0.430 \\
Text+Chem T5 &0.625 & 0.542 & 0.682 & 0.543 & 0.622 &0.648 \\
BioT5 & 0.635 & 0.556 & 0.692 & 0.559 & 0.633 & 0.656 \\
MolCA & 0.639 & 0.555 & 0.697 & 0.558 & 0.636 & 0.669 \\
\midrule
\rowcolor[RGB]{234, 238, 234} \multicolumn{7}{l}{\textit{Retrieval Based LLMs}} \\
GPT-3.5-turbo (10-shot MolReGPT) &0.565 &0.482 &0.623 &0.450 &0.543 &0.585 \\
GPT-4-0314 (10-shot MolReGPT) &0.607 &0.525 &0.634 &0.476 &0.562 &0.610 \\
\midrule
\rowcolor[RGB]{234, 238, 234} \multicolumn{7}{l}{\textit{LLM-based Generalist Models}} \\
GPT-3.5-turbo (zero-shot) & 0.103 &0.050 &0.261 &0.088 &0.204 &0.161 \\
BioMedGPT-10B & 0.234 &0.141  &0.386  &0.206  &0.332  &0.308  \\
Mol-Instructions-7B & 0.249 &0.171  &0.331  &0.203  &0.289  &0.271  \\
InstructMol-G-6.9B & 0.466	&0.365	&0.547	&0.365	&0.479	&0.491	\\
InstructMol-GS-6.9B & 0.475 & 0.371 & 0.566 & 0.394 &0.502 &0.509 \\
\midrule
\textbf{\method} & \textbf{0.666} & \textbf{0.591} & \textbf{0.710} & \textbf{0.584} & \textbf{0.650} & \textbf{0.681} \\
\bottomrule
\end{tabular}
}}
\label{tab:mol2text}
\vspace{-0.3cm}
\end{table*}

\begin{table*}[t]
\caption{\footnotesize{Performance comparison on description-guided molecule design task on ChEBI-20~\cite{molt5} dataset.
The ground truth Text2Mol~\cite{text2mol} score is $0.609$.
}}
\resizebox{\textwidth}{!}{
\centering
\begin{tabular}{lccccccccc}
\toprule
\textsc{Model} & \textsc{BLEU}$\uparrow$ & \textsc{Exact}$\uparrow$ & \textsc{Levenshtein}$\downarrow$ & \textsc{MACCS FTS}$\uparrow$ & \textsc{RDK FTS}$\uparrow$ & \textsc{Morgan FTS}$\uparrow$ & \textsc{FCD}$\downarrow$ & \textsc{Text2Mol}$\uparrow$ & \textsc{Validity}$\uparrow$ \\
\midrule
\rowcolor[RGB]{234, 238, 234} \multicolumn{10}{l}{\textit{Single-task Specialist Models}} \\
Transformer & 0.499 & 0.000 & 57.660 & 0.480 & 0.320 & 0.217 & 11.32 & 0.277 & 0.906 \\
T5-base & 0.762 & 0.069 & 24.950 & 0.731 &  0.605 & 0.545 & 2.48 & 0.499 & 0.660  \\
MolT5-base & 0.769 & 0.081 & 24.458 & 0.721 &  0.588 & 0.529 & 2.18 & 0.496 & 0.772 \\
MoMu-base & 0.815 & 0.183 & 20.520 & 0.847 & 0.737 & 0.678 & - & 0.580 & 0.863 \\
MolFM-base & 0.822 & 0.210 & 19.445 & 0.854 & 0.758 & 0.697 & - & 0.583 & 0.892 \\
GIT-Mol & 0.756 & 0.051 & 26.315 & 0.738 & 0.582 & 0.519 & - & - & 0.928 \\
MolXPT & - & 0.215 & - & 0.859 & 0.757 & 0.667 & 0.45 & 0.578 & 0.983 \\
BioT5 & 0.867 & 0.413 & 15.097 & 0.886 & 0.801 & 0.734 & 0.43 & 0.576 & \textbf{1.000} \\
\rowcolor[RGB]{234, 238, 234} \multicolumn{10}{l}{\textit{Retrieval-based LLMs}} \\
Llama2-7B (2-shot MolReGPT) & 0.693 & 0.022 & 36.77 & 0.808 & 0.717 & 0.609 & 4.90 & 0.149 & 0.761 \\
GPT-3.5-turbo (10-shot MolReGPT) & 0.790 & 0.139 & 24.91 & 0.847 & 0.708 & 0.624 & 0.57 & 0.571 & 0.887\\
GPT-4-0314 (10-shot MolReGPT) & 0.857 & 0.280 & 17.14 & 0.903 & 0.805 & 0.739 & 0.41 & \textbf{0.593} & 0.899 \\\midrule
\rowcolor[RGB]{234, 238, 234} \multicolumn{10}{l}{\textit{LLM-based Generalist Models}} \\
Llama2-7B (0-shot) & 0.104 & 0.000 & 84.18 & 0.243 & 0.119 & 0.089 & 42.01 & 0.148 & 0.631\\
GPT-3.5-turbo (0-shot) & 0.489 & 0.019 & 52.13 & 0.705 & 0.462 & 0.367 & 2.05 & 0.479 & 0.802 \\
\midrule
\textbf{\method} & \textbf{0.872} & \textbf{0.522} & \textbf{12.776} & \textbf{0.907} & \textbf{0.835} & \textbf{0.779} & \textbf{0.353} & 0.579 & \textbf{1.000} \\
\bottomrule
\end{tabular}
}
\label{tab:text2mol}
\vspace{-0.5cm}
\end{table*}

\begin{table}[t!]
\centering
\caption{\footnotesize{Performance (accuracy) comparison on PEER benchmark (\textbf{Best}, \underline{Second Best}). 
* means linear probing.}}
\resizebox{\linewidth}{!}{
\small
\begin{tabular}{lcccc}
\toprule
\textsc{Model} & \textsc{Solubility} & \textsc{Localization} & \textsc{Yeast} & \textsc{Human}\\
\midrule
\rowcolor[RGB]{234, 238, 234} \multicolumn{5}{l}{\textit{Single-task Specialist Models}} \\
DDE & 59.77 $\pm$ 1.21 & 77.43 $\pm$ 0.42 & 55.83 $\pm$ 3.13 & 62.77 $\pm$ 2.30 \\
Moran & 57.73 $\pm$ 1.33 & 55.63 $\pm$ 0.85 & 53.00 $\pm$ 0.50 & 54.67 $\pm$ 4.43\\
LSTM & 70.18 $\pm$ 0.63 & 88.11 $\pm$ 0.14 & 53.62 $\pm$ 2.72 & 63.75 $\pm$ 5.12\\
Transformer & 70.12 $\pm$ 0.31 & 75.74 $\pm$ 0.74 & 54.12 $\pm$ 1.27 & 59.58 $\pm$ 2.09\\
CNN & 64.43 $\pm$ 0.25 & 82.67 $\pm$ 0.32 & 55.07 $\pm$ 0.02 & 62.60 $\pm$ 1.67\\
ResNet & 67.33 $\pm$ 1.46 & 78.99 $\pm$ 4.41 & 48.91 $\pm$ 1.78 & 68.61 $\pm$ 3.78\\
ProtBert & 68.15 $\pm$ 0.92 & 91.32 $\pm$ 0.89 & 63.72 $\pm$ 2.80 & 77.32 $\pm$ 1.10\\
ProtBert* & 59.17 $\pm$ 0.21 & 81.54 $\pm$ 0.09 & 53.87 $\pm$ 0.38 & 83.61 $\pm$ 1.34\\
ESM-1B & \underline{70.23 $\pm$ 0.75} & \textbf{92.40 $\pm$ 0.35} & 57.00 $\pm$ 6.38 & 78.17 $\pm$ 2.91\\
ESM-1B* & 67.02 $\pm$ 0.40 & 91.61 $\pm$ 0.10 & \textbf{66.07 $\pm$ 0.58} & \textbf{88.06 $\pm$ 0.24} \\
BioT5 & \textbf{74.65 $\pm$ 0.49} & \underline{91.69 $\pm$ 0.05} & \underline{64.89 $\pm$ 0.43} & \underline{86.22 $\pm$ 0.53}\\
\midrule
\rowcolor[RGB]{234, 238, 234} \multicolumn{5}{l}{\textit{Multi-task Generalist Models}} \\
CNN & 70.63 $\pm$ 0.34 & 82.67 $\pm$ 0.72 & 54.50 $\pm$ 1.61 & 69.03 $\pm$ 2.68\\
Transformer & 70.03 $\pm$ 0.42 & 76.27 $\pm$ 0.57 & 54.00 $\pm$ 1.17 & 67.33 $\pm$ 2.68\\
ESM-1B & \underline{70.46 $\pm$ 0.16} & \textbf{92.50 $\pm$ 0.26} & \underline{64.76 $\pm$ 1.42} & \underline{83.00 $\pm$ 0.88}\\
\midrule
\textbf{\method} & \textbf{74.37 $\pm$ 0.19} & \underline{90.41 $\pm$ 0.07} & \textbf{66.16 $\pm$ 0.43} & \textbf{85.09 $\pm$ 0.40} \\
\bottomrule
\end{tabular}
}
\label{tab:peer}
\vspace{-0.75cm}
\end{table}

\subsubsection{Chemical Reaction-related Tasks}
In computational chemistry, tasks related to chemical reactions are of vital importance as they can speed up development processes.
Following~\citealp{instructmol}, we focus on 3 such tasks: reagent prediction, forward reaction prediction, and retrosynthesis.

\noindent{\textbf{Results.}} The main results are presented in Table~\ref{tab:chemical_reaction} and full results are in Table~\ref{tab:chemical_reaction_full}.
While LLMs have been exposed to some molecular data during pre-training, their direct zero-shot testing on chemical reaction-related tasks demonstrated extremely poor performance. 
Mol-Instructions~\citep{mol-instructions} conducts multi-task instruction tuning based on Llama~\citep{llama} with molecule-oriented tasks. 
InstructMol~\citep{instructmol} introduces a molecule graph encoder to encode 2D molecular graph information for Vicuna~\citep{vicuna}.
Our \method{} follows the same training setting with Mol-Instructions~\citep{mol-instructions} and shows superior performance across almost all metrics on chemical reaction-related tasks. 
This outcome demonstrates the effectiveness of joint pre-training on both molecular and textual data. 
\begin{table*}[!t]
\vspace{-1.3cm}
\centering
\begin{minipage}{0.66\textwidth}
\centering
\caption{\footnotesize{
Ablation of IUPAC and additional data on molecule description generation task with single task setting.
B-2 stands for BLEU-2 and R-1 is ROUGE-1.
}}
\small
\resizebox{\textwidth}{!}{
\begin{tabular}{p{2.5cm}cccccc}
\toprule
\textsc{Model}
&\textsc{B-2}$\uparrow$  & \textsc{B-4}$\uparrow$  & \textsc{R-1}$\uparrow$  & \textsc{R-2}$\uparrow$  & \textsc{R-L}$\uparrow$ & \textsc{METEOR}$\uparrow$ \\
\midrule
\method (single task) & 0.671 & 0.597 & 0.715 & 0.590 & 0.655 & 0.687 \\
\method (single task) wo IUPAC & 0.661 & 0.584 & 0.706 & 0.578 & 0.647 & 0.677 \\
\method (single task) wo additional data & 0.666 & 0.591 & 0.711 & 0.586 & 0.651 & 0.681 \\
\midrule
\method & 0.666 & 0.591 & 0.710 & 0.584 & 0.650 & 0.681 \\
\bottomrule
\end{tabular}
}
\label{tab:mol2text_abl}
\end{minipage}
\hfill
\begin{minipage}{0.3\textwidth}
	\centering
        \vspace{-0.4cm}
	\includegraphics[width=\textwidth]{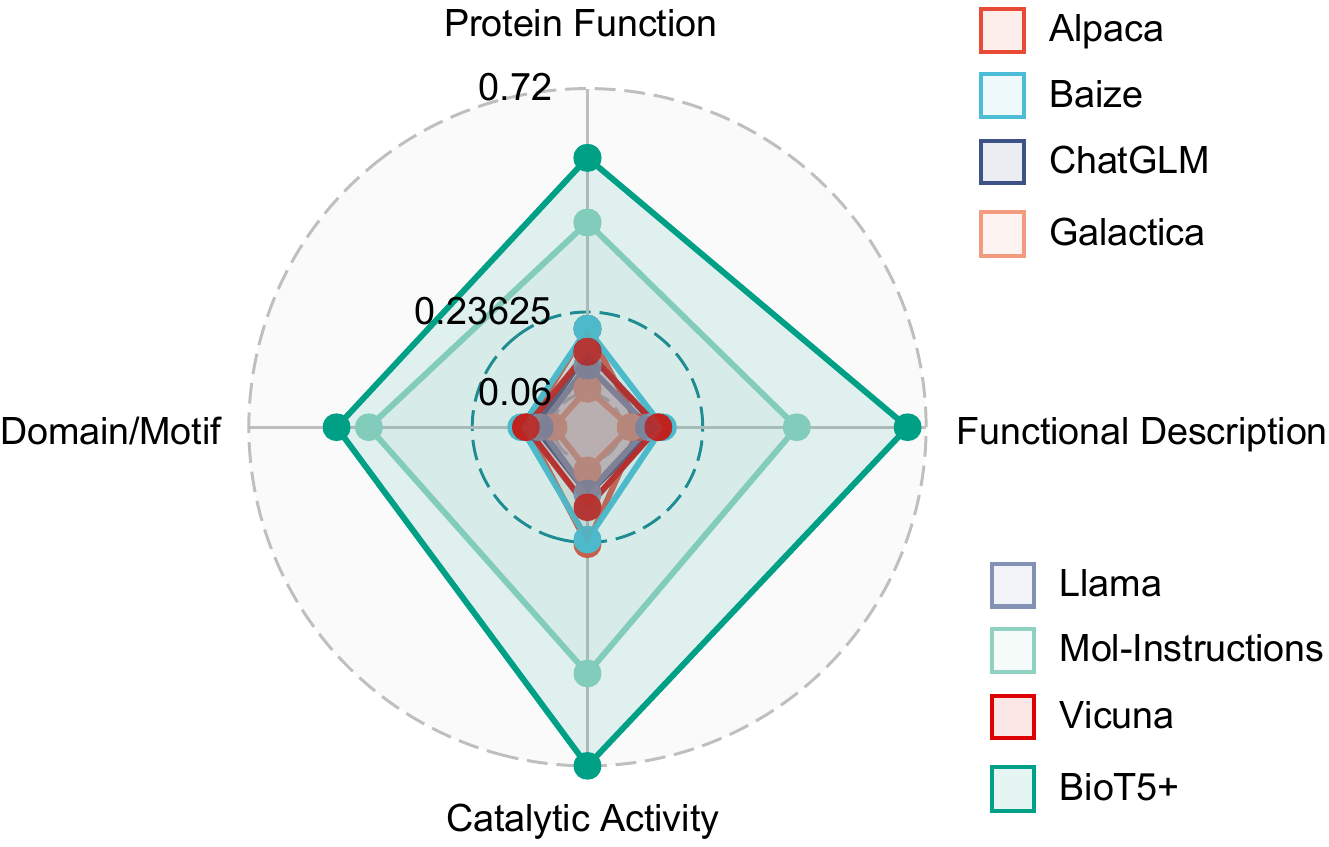}
	\captionof{figure}{\footnotesize{Performance (ROUGE-L) comparison on protein description generation tasks.}}
	\label{fig:molinst_pro_understand}
\end{minipage}
\vspace{-0.5cm}
\end{table*}
\begin{table}[t!]
\centering
\caption{\footnotesize{
Performance (AUROC) comparison on the 3 DTI datasets(\textbf{Best}, \underline{Second Best}).
}}
\resizebox{\linewidth}{!}{
\begin{tabular}{c c c c}
\toprule
\textsc{Method} & \textsc{BioSNAP} & \textsc{Human} & \textsc{BindingDB} \\
\midrule
\rowcolor[RGB]{234, 238, 234} \multicolumn{4}{l}{\textit{Single-task Specialist Models}} \\
SVM             & 0.862$\pm$0.007     & 0.940$\pm$0.006  & 0.939$\pm$0.001 \\
RF        & 0.860$\pm$0.005       & 0.952$\pm$0.011   & 0.942$\pm$0.011  \\
DeepConv-DTI           & 0.886$\pm$0.006  & 0.980$\pm$0.002  & 0.945$\pm$0.002 \\
GraphDTA      & 0.887$\pm$0.008  & 0.981$\pm$0.001  & 0.951$\pm$0.002 \\
MolTrans & 0.895$\pm$0.004       & 0.980$\pm$0.002  & 0.952$\pm$0.002 \\
DrugBAN & \underline{0.903$\pm$0.005}  & \underline{0.982$\pm$0.002}  & \underline{0.960$\pm$0.001}\\
BioT5 & 0.937$\pm$0.001  & \textbf{0.989$\pm$0.001}  & 0.963$\pm$0.001 \\
\midrule
\rowcolor[RGB]{234, 238, 234} \multicolumn{4}{l}{\textit{Multi-task Generalist Models}} \\
\method & \textbf{0.939$\pm$0.001} & 0.987$\pm$0.001  & \textbf{0.964$\pm$0.001} \\
\bottomrule
\end{tabular}
}
\label{tab:dti}
\vspace{-0.3cm}
\end{table}
\begin{table}
\centering
\small
\caption{\footnotesize
{Ablation of default T5 tokenizer and character-based tokenizer (\method) on QM9 dataset.
}
}
\resizebox{\linewidth}{!}{
\begin{tabular}{lccccc}
    \toprule
    \textsc{Method} &\textsc{HOMO} $\downarrow$ &\textsc{LUMO} $\downarrow$  &$\Delta{\epsilon}$ $\downarrow$ &\textsc{Avg} $\downarrow$\\
    \midrule
    T5 default tokenizer & 0.0024 & 0.0026 & 0.0032 & 0.0027 \\
    \method & 0.0022 & 0.0024 & 0.0028 & 0.0025 \\
    \bottomrule
\end{tabular}
}
\label{tab:abl_num}
\vspace{-0.6cm}
\end{table}

\subsubsection{Molecule Description Generation}
The objective of molecule description generation is to generate a detailed and informative description for a given molecule. 
To be consistent with \method{} pre-training, the input here also consists of molecule SELFIES with IUPAC. 
Unlike molecule property prediction, which often focuses on specific attributes, molecule description generation involves interpreting and conveying a comprehensive narrative of the molecule. This narrative encompasses not only its molecular composition and properties but also its potential applications and roles, as derived from the integration of SELFIES representations and IUPAC names.
We use the same evaluation metrics as~\citealp{mol-instructions}.

\noindent{\textbf{Results.}}
As presented in Table~\ref{tab:mol2text}, our \method{} outperforms all compared single-task specialist, retrieval-based LLMs, and multi-task generalist baselines. This superior performance can be attributed to the comprehensive learning during the pre-training of \method{}.
The model has effectively assimilated a multi-dimensional and rich textual description of molecules. 

\subsubsection{Description-guided Molecule Design}
Description-guided molecule design is essentially the inverse task of molecule description generation, which requires generating a molecule based on a provided textual description.
In \method{} setting, we do not include IUPAC name in the textual description of the molecule to prevent the model from learning a simplistic mapping from IUPAC name to its SELFIES representation, thereby ensuring the model does not overlook the other descriptive elements provided in the text. 

\noindent{\textbf{Results.}} Table~\ref{tab:text2mol} presents the results for description-guided molecule design task.
Our \method{} surpasses all the compared baselines. This achievement underscores the efficacy of \method{} pre-training, where the model has acquired a profound understanding of molecular knowledge. 

\subsection{Protein-oriented Tasks}
\subsubsection{Protein Description Generation}
\label{sec:protein_understand_generation}
The task of protein description generation involves deriving relevant textual information from a given protein sequence. 
Following~\citealp{mol-instructions}, we mainly focus on 4 related generation tasks: protein function generation, catalytic activity generation, domain/motif generation, and functional description generation.

\noindent{\textbf{Results.}} As shown in the Figure~\ref{fig:molinst_pro_understand}, \method{} surpasses all compared baselines across 4 tasks.
This result highlights the advanced ability of \method{} to interpret complex protein sequences into meaningful textual information, indicating that \method{} gains a comprehensive understanding of protein structures and functions through pre-training. 

\subsubsection{Protein Property Prediction}
Protein property prediction task involves predicting specific properties of proteins, such as solubility, structure, or function, based on their amino acid sequences. 
Following BioT5~\cite{biot5}, we focus on 2 protein property prediction tasks from PEER~\citep{peer} benchmark, which is specifically designed for protein sequence understanding:
(1) Solubility prediction: whether the input protein is soluble or not.
(2) Localization prediction: either the input protein is ``membrane-bound'' or ``soluble''.
Both of these tasks are binary classification tasks and the model needs to generate a ``yes'' or ``no'' prediction.
The results are summarized together with the ones in the following Section~\ref{sec:protein_interaction}.
\subsubsection{Protein-related Interaction Prediction}
\label{sec:protein_interaction}
In drug discovery, the prediction of interactions between bio-entities is very important, from which Protein-Protein Interaction (PPI) and Drug-Target Interactions (DTI) are two key examples.
These two tasks are essential for understanding biological processes and identifying potential therapeutic targets. 
To facilitate this, we follow~\citealp{biot5} to incorporate 2 PPI dataset from PEER~\cite{peer} benchmark including Yeast and Human, and 3 DTI datasets including BioSNAP~\citep{zitnik2018biosnap}, Human~\citep{liu2015improving,chen2020transformercpi}, and BindingDB~\citep{liu2007bindingdb}.

\noindent{\textbf{Results.}}
As shown in Table~\ref{tab:peer}, in the PEER benchmark, our \method{} demonstrates exceptional performance, surpassing other multi-task models in 3 out of the 4 tasks and achieving results comparable to single-task specialist models. 
Notably, in the Yeast PPI prediction task, \method{} exceeded the performance of all baseline models. This is particularly significant considering that the baseline ESM-1b~\citep{esm} was specifically pre-trained on a vast array of protein sequences and possesses more than double the number of parameters compared to \method.
Furthermore, \method{} also showed superior performance in DTI tasks as in Table~\ref{tab:dti} (full results in Table~\ref{tab:dti_full}), consistently outperforming other methods on the BioSNAP and BindingDB datasets. It is noteworthy that many baseline methods involved specialized designs for molecule and protein encoders.
These results underscore the effectiveness of the joint pre-training of \method{} on bio-text, molecules, and proteins. 
This comprehensive understanding is evident in the ability of \method{} to accurately predict protein properties, interactions, and drug-target interactions, making it a valuable tool in the field of computational biology.

\subsubsection{Description-guided Protein Design}
For description-guided protein design, the model needs to generate protein amino acid sequences based on specific design requirements, such as protein structures and functions.
Due to the absence of a well-established benchmark for this task, we present in Appendix Table~\ref{appendix:text2pro_cases} a selection of test cases along with their corresponding sequence similarity scores to provide a direct comparison between our model and existing models like Galactica~\cite{galactica} and Mol-Instructions~\cite{mol-instructions}.

\subsection{Ablation Studies}
\label{sec:ablation}
In this section, we conduct ablation studies to investigate the effectiveness of our design in \method{}.
Specifically, we focus on the following 3 scenarios:
\textbf{(1) Do not incorporate IUPAC name of the molecule.} As shown in Table~\ref{tab:mol2text_abl}, removing the IUPAC name results in a noticeable performance drop in the molecule description generation task. This decline highlights the significant role that IUPAC name plays in tasks related to molecular understanding.
\textbf{(2) Do not add PubMed Central and bioRxiv data in pre-training.} Results in Table~\ref{tab:mol2text_abl} and Table~\ref{tab:text2mol_abl} indicate that these two datasets play a crucial role in enhancing molecular understanding. The omission of them leads to a slight but noticeable decrease in performance on molecule description generation and description-guided molecule design tasks.
\textbf{(3) Use T5 default tokenizer for numbers instead of character-based tokenizer.} The results in Table~\ref{tab:abl_num} demonstrate that the character-based approach for tokenizing numbers is more effective than the default T5 tokenizer on regression tasks.
We also conduct an ablation study to further contrast single-task and multi-task tuning strategies in Appendix Section~\ref{appsec:ablation}.

\section{Conclusions}
\method{}, as an advanced iteration of the BioT5 framework, represents a significant stride in computational biology and drug discovery. 
By integrating IUPAC names, expanding bio-text and molecular data sources, employing multi-task instruction tuning, and incorporating an advanced numerical tokenization technique, \method{} has successfully bridged the gap between molecular representations and their textual descriptions. 
The enhanced understanding of molecular structures and its ability to process complex biological data have been demonstrated across a wide range of tasks, with \method{} achieving state-of-the-art performance in most of them. 
This success highlights the potential of \method{} as a versatile and powerful tool in understanding and analyzing biological entities.

\section{Limitations}
Despite the significant advancements achieved by \method{}, there remain certain limitations that need to be addressed in future work. 
Firstly, the model faces challenges in generalizing across various biological tasks, a problem that is distinct from common NLP settings. The intricate and unique nature of each biological task makes it difficult to develop a one-size-fits-all solution, highlighting the need for more specialized approaches within this domain.
Secondly, the current scale of \method{} is somewhat limited and cannot comprehend and integrate information from other modalities, such as images, restricting its applicability in multi-modal biological data analysis. \method{} is not equipped to function as a universal chatbot or to answer queries spanning general domain questions outside the specific scope of biology. 
This constraint highlights the need for developing larger, more versatile models capable of handling a wider range of data types and answering a broader array of questions in both biological and general domains.

\section{Ethical Considerations}
While \method{} presents a significant advancement, its capabilities, particularly in generating molecules based on textual descriptions and predicting chemical reaction products, raise important ethical considerations. 
One of the concerns is the potential misuse of this technology to generate harmful or dangerous molecules, which could pose risks to public safety and environmental health.
Moreover, the ability of \method{} to predict and generate novel molecules may also lead to issues surrounding intellectual property rights and patenting. The ease with which new compounds can be designed and synthesized using AI-driven methods could potentially disrupt traditional research practices and raise questions about the ownership of these discoveries.

\section{Acknowledgements}
This work was supported by the National Natural Science Foundation of China (NSFC Grant No. 62122089), Beijing Outstanding Young Scientist Program NO. BJJWZYJH012019100020098, and Intelligent Social Governance Platform, Major Innovation \& Planning Interdisciplinary Platform for the "Double-First Class" Initiative, Renmin University of China, the Fundamental Research Funds for the Central Universities, and the Research Funds of Renmin University of China.
Qizhi Pei is supported by the Outstanding Innovative Talents Cultivation Funded Programs 2023 of Renmin University of China.

\bibliography{custom}

\begin{thebibliography}{88}
\expandafter\ifx\csname natexlab\endcsname\relax\def\natexlab#1{#1}\fi

\bibitem[{Abdine et~al.(2023)Abdine, Chatzianastasis, Bouyioukos, and Vazirgiannis}]{prottext}
Hadi Abdine, Michail Chatzianastasis, Costas Bouyioukos, and Michalis Vazirgiannis. 2023.
\newblock \href {https://openreview.net/forum?id=XPDudqlrEW} {Prot2text: Multimodal protein's function generation with {GNN}s and transformers}.
\newblock In \emph{NeurIPS 2023 AI for Science Workshop}.

\bibitem[{AI4Science and Quantum(2023)}]{ai4science2023impact}
Microsoft~Research AI4Science and Microsoft~Azure Quantum. 2023.
\newblock The impact of large language models on scientific discovery: a preliminary study using gpt-4.
\newblock \emph{arXiv preprint arXiv:2311.07361}.

\bibitem[{Bai et~al.(2023)Bai, Miljkovi{\'c}, John, and Lu}]{drugban}
Peizhen Bai, Filip Miljkovi{\'c}, Bino John, and Haiping Lu. 2023.
\newblock Interpretable bilinear attention network with domain adaptation improves drug--target prediction.
\newblock \emph{Nature Machine Intelligence}, 5(2):126--136.

\bibitem[{Banerjee and Lavie(2005)}]{meteor}
Satanjeev Banerjee and Alon Lavie. 2005.
\newblock \href {https://aclanthology.org/W05-0909/} {{METEOR:} an automatic metric for {MT} evaluation with improved correlation with human judgments}.
\newblock In \emph{Proceedings of the Workshop on Intrinsic and Extrinsic Evaluation Measures for Machine Translation and/or Summarization@ACL 2005, Ann Arbor, Michigan, USA, June 29, 2005}, pages 65--72. Association for Computational Linguistics.

\bibitem[{Brown et~al.(2020)Brown, Mann, Ryder, Subbiah, Kaplan, Dhariwal, Neelakantan, Shyam, Sastry, Askell et~al.}]{gpt3}
Tom Brown, Benjamin Mann, Nick Ryder, Melanie Subbiah, Jared~D Kaplan, Prafulla Dhariwal, Arvind Neelakantan, Pranav Shyam, Girish Sastry, Amanda Askell, et~al. 2020.
\newblock Language models are few-shot learners.
\newblock \emph{Advances in neural information processing systems}, 33:1877--1901.

\bibitem[{Canese and Weis(2013)}]{pubmed1.0}
Kathi Canese and Sarah Weis. 2013.
\newblock Pubmed: the bibliographic database.
\newblock \emph{The NCBI handbook}, 2(1).

\bibitem[{Cao et~al.(2023)Cao, Liu, Lu, Yao, and Li}]{instructmol}
He~Cao, Zijing Liu, Xingyu Lu, Yuan Yao, and Yu~Li. 2023.
\newblock \href {https://doi.org/10.48550/ARXIV.2311.16208} {Instructmol: Multi-modal integration for building a versatile and reliable molecular assistant in drug discovery}.
\newblock \emph{CoRR}, abs/2311.16208.

\bibitem[{Chen et~al.(2020)Chen, Tan, Wang, Zhong, Liu, Yang, Luo, Chen, Jiang, and Zheng}]{chen2020transformercpi}
Lifan Chen, Xiaoqin Tan, Dingyan Wang, Feisheng Zhong, Xiaohong Liu, Tianbiao Yang, Xiaomin Luo, Kaixian Chen, Hualiang Jiang, and Mingyue Zheng. 2020.
\newblock Transformercpi: improving compound--protein interaction prediction by sequence-based deep learning with self-attention mechanism and label reversal experiments.
\newblock \emph{Bioinformatics}, 36(16):4406--4414.

\bibitem[{Chiang et~al.(2023)Chiang, Li, Lin, Sheng, Wu, Zhang, Zheng, Zhuang, Zhuang, Gonzalez, Stoica, and Xing}]{vicuna}
Wei-Lin Chiang, Zhuohan Li, Zi~Lin, Ying Sheng, Zhanghao Wu, Hao Zhang, Lianmin Zheng, Siyuan Zhuang, Yonghao Zhuang, Joseph~E. Gonzalez, Ion Stoica, and Eric~P. Xing. 2023.
\newblock \href {https://lmsys.org/blog/2023-03-30-vicuna/} {Vicuna: An open-source chatbot impressing gpt-4 with 90\%* chatgpt quality}.

\bibitem[{Chithrananda et~al.(2020)Chithrananda, Grand, and Ramsundar}]{chemberta}
Seyone Chithrananda, Gabriel Grand, and Bharath Ramsundar. 2020.
\newblock Chemberta: Large-scale self-supervised pretraining for molecular property prediction.
\newblock \emph{arXiv preprint arXiv:2010.09885}.

\bibitem[{Christofidellis et~al.(2023)Christofidellis, Giannone, Born, Winther, Laino, and Manica}]{text+chemt5}
Dimitrios Christofidellis, Giorgio Giannone, Jannis Born, Ole Winther, Teodoro Laino, and Matteo Manica. 2023.
\newblock \href {https://proceedings.mlr.press/v202/christofidellis23a.html} {Unifying molecular and textual representations via multi-task language modelling}.
\newblock In \emph{International Conference on Machine Learning, {ICML} 2023, 23-29 July 2023, Honolulu, Hawaii, {USA}}, volume 202 of \emph{Proceedings of Machine Learning Research}, pages 6140--6157. {PMLR}.

\bibitem[{Cortes and Vapnik(1995)}]{cortes1995support}
Corinna Cortes and Vladimir Vapnik. 1995.
\newblock Support-vector networks.
\newblock \emph{Machine learning}, 20(3):273--297.

\bibitem[{Dara et~al.(2022)Dara, Dhamercherla, Jadav, Babu, and Ahsan}]{ml_drug_discovery}
Suresh Dara, Swetha Dhamercherla, Surender~Singh Jadav, Ch~Madhu Babu, and Mohamed~Jawed Ahsan. 2022.
\newblock \href {https://doi.org/10.1007/S10462-021-10058-4} {Machine learning in drug discovery: {A} review}.
\newblock \emph{Artif. Intell. Rev.}, 55(3):1947--1999.

\bibitem[{Durant et~al.(2002)Durant, Leland, Henry, and Nourse}]{durant2002reoptimization}
Joseph~L Durant, Burton~A Leland, Douglas~R Henry, and James~G Nourse. 2002.
\newblock Reoptimization of mdl keys for use in drug discovery.
\newblock \emph{Journal of chemical information and computer sciences}, 42(6):1273--1280.

\bibitem[{Edwards et~al.(2022)Edwards, Lai, Ros, Honke, Cho, and Ji}]{molt5}
Carl Edwards, Tuan~Manh Lai, Kevin Ros, Garrett Honke, Kyunghyun Cho, and Heng Ji. 2022.
\newblock \href {https://aclanthology.org/2022.emnlp-main.26} {Translation between molecules and natural language}.
\newblock In \emph{Proceedings of the 2022 Conference on Empirical Methods in Natural Language Processing, {EMNLP} 2022, Abu Dhabi, United Arab Emirates, December 7-11, 2022}, pages 375--413. Association for Computational Linguistics.

\bibitem[{Edwards et~al.(2021)Edwards, Zhai, and Ji}]{text2mol}
Carl Edwards, ChengXiang Zhai, and Heng Ji. 2021.
\newblock \href {https://doi.org/10.18653/v1/2021.emnlp-main.47} {Text2mol: Cross-modal molecule retrieval with natural language queries}.
\newblock In \emph{Proceedings of the 2021 Conference on Empirical Methods in Natural Language Processing, {EMNLP} 2021, Virtual Event / Punta Cana, Dominican Republic, 7-11 November, 2021}, pages 595--607. Association for Computational Linguistics.

\bibitem[{Elnaggar et~al.(2021)Elnaggar, Heinzinger, Dallago, Rehawi, Wang, Jones, Gibbs, Feher, Angerer, Steinegger et~al.}]{elnaggar2021prottrans}
Ahmed Elnaggar, Michael Heinzinger, Christian Dallago, Ghalia Rehawi, Yu~Wang, Llion Jones, Tom Gibbs, Tamas Feher, Christoph Angerer, Martin Steinegger, et~al. 2021.
\newblock Prottrans: Toward understanding the language of life through self-supervised learning.
\newblock \emph{IEEE transactions on pattern analysis and machine intelligence}, 44(10):7112--7127.

\bibitem[{et.al.(2023)}]{chatglm}
Aohan~Zeng et.al. 2023.
\newblock \href {https://openreview.net/forum?id=-Aw0rrrPUF} {{GLM}-130b: An open bilingual pre-trained model}.
\newblock In \emph{The Eleventh International Conference on Learning Representations (ICLR)}.

\bibitem[{Fang et~al.(2022)Fang, Liu, Lei, He, Zhang, Zhou, Wang, Wu, and Wang}]{gem}
Xiaomin Fang, Lihang Liu, Jieqiong Lei, Donglong He, Shanzhuo Zhang, Jingbo Zhou, Fan Wang, Hua Wu, and Haifeng Wang. 2022.
\newblock Geometry-enhanced molecular representation learning for property prediction.
\newblock \emph{Nature Machine Intelligence}, 4(2):127--134.

\bibitem[{Fang et~al.(2023)Fang, Liang, Zhang, Liu, Huang, Chen, Fan, and Chen}]{mol-instructions}
Yin Fang, Xiaozhuan Liang, Ningyu Zhang, Kangwei Liu, Rui Huang, Zhuo Chen, Xiaohui Fan, and Huajun Chen. 2023.
\newblock \href {https://doi.org/10.48550/ARXIV.2306.08018} {Mol-instructions: {A} large-scale biomolecular instruction dataset for large language models}.
\newblock \emph{CoRR}, abs/2306.08018.

\bibitem[{Feng and Zhang(2000)}]{feng2000prediction}
Zhi-Ping Feng and Chun-Ting Zhang. 2000.
\newblock Prediction of membrane protein types based on the hydrophobic index of amino acids.
\newblock \emph{Journal of protein chemistry}, 19:269--275.

\bibitem[{He et~al.(2016)He, Zhang, Ren, and Sun}]{he2016deep}
Kaiming He, Xiangyu Zhang, Shaoqing Ren, and Jian Sun. 2016.
\newblock Deep residual learning for image recognition.
\newblock In \emph{Proceedings of the IEEE conference on computer vision and pattern recognition}, pages 770--778.

\bibitem[{Ho(1995)}]{ho1995random}
Tin~Kam Ho. 1995.
\newblock Random decision forests.
\newblock In \emph{Proceedings of 3rd International Conference on Document Analysis and Recognition}, volume~1, pages 278--282.

\bibitem[{Hochreiter and Schmidhuber(1997)}]{hochreiter1997long}
Sepp Hochreiter and J{\"u}rgen Schmidhuber. 1997.
\newblock Long short-term memory.
\newblock \emph{Neural computation}, 9(8):1735--1780.

\bibitem[{Hu et~al.(2022)Hu, Shen, Wallis, Allen{-}Zhu, Li, Wang, Wang, and Chen}]{lora}
Edward~J. Hu, Yelong Shen, Phillip Wallis, Zeyuan Allen{-}Zhu, Yuanzhi Li, Shean Wang, Lu~Wang, and Weizhu Chen. 2022.
\newblock \href {https://openreview.net/forum?id=nZeVKeeFYf9} {Lora: Low-rank adaptation of large language models}.
\newblock In \emph{The Tenth International Conference on Learning Representations, {ICLR} 2022, Virtual Event, April 25-29, 2022}. OpenReview.net.

\bibitem[{Huang et~al.(2021)Huang, Xiao, Glass, and Sun}]{Huang2021MolTransMI}
Kexin Huang, Cao Xiao, Lucas Glass, and Jimeng Sun. 2021.
\newblock {MolTrans}: Molecular interaction transformer for drug–target interaction prediction.
\newblock \emph{Bioinformatics}, 37:830 -- 836.

\bibitem[{Irwin et~al.(2020)Irwin, Tang, Young, Dandarchuluun, Wong, Khurelbaatar, Moroz, Mayfield, and Sayle}]{irwin2020zinc20}
John~J Irwin, Khanh~G Tang, Jennifer Young, Chinzorig Dandarchuluun, Benjamin~R Wong, Munkhzul Khurelbaatar, Yurii~S Moroz, John Mayfield, and Roger~A Sayle. 2020.
\newblock Zinc20—a free ultralarge-scale chemical database for ligand discovery.
\newblock \emph{Journal of chemical information and modeling}, 60(12):6065--6073.

\bibitem[{Kim et~al.(2019)Kim, Chen, Cheng, Gindulyte, He, He, Li, Shoemaker, Thiessen, Yu et~al.}]{kim2019pubchem}
Sunghwan Kim, Jie Chen, Tiejun Cheng, Asta Gindulyte, Jia He, Siqian He, Qingliang Li, Benjamin~A Shoemaker, Paul~A Thiessen, Bo~Yu, et~al. 2019.
\newblock Pubchem 2019 update: improved access to chemical data.
\newblock \emph{Nucleic acids research}, 47(D1):D1102--D1109.

\bibitem[{Klinger et~al.(2008)Klinger, Kol{\'{a}}rik, Fluck, Hofmann{-}Apitius, and Friedrich}]{iupac_detect}
Roman Klinger, Corinna Kol{\'{a}}rik, Juliane Fluck, Martin Hofmann{-}Apitius, and Christoph~M. Friedrich. 2008.
\newblock \href {https://doi.org/10.1093/BIOINFORMATICS/BTN181} {Detection of {IUPAC} and iupac-like chemical names}.
\newblock In \emph{Proceedings 16th International Conference on Intelligent Systems for Molecular Biology (ISMB), Toronto, Canada, July 19-23, 2008}, pages 268--276.

\bibitem[{Krenn et~al.(2020)Krenn, H{\"a}se, Nigam, Friederich, and Aspuru-Guzik}]{krenn2020self}
Mario Krenn, Florian H{\"a}se, AkshatKumar Nigam, Pascal Friederich, and Alan Aspuru-Guzik. 2020.
\newblock Self-referencing embedded strings (selfies): A 100\% robust molecular string representation.
\newblock \emph{Machine Learning: Science and Technology}, 1(4):045024.

\bibitem[{Kudo and Richardson(2018)}]{DBLP:conf/emnlp/KudoR18}
Taku Kudo and John Richardson. 2018.
\newblock \href {https://doi.org/10.18653/v1/d18-2012} {Sentencepiece: {A} simple and language independent subword tokenizer and detokenizer for neural text processing}.
\newblock In \emph{Proceedings of the 2018 Conference on Empirical Methods in Natural Language Processing, {EMNLP} 2018: System Demonstrations, Brussels, Belgium, October 31 - November 4, 2018}, pages 66--71. Association for Computational Linguistics.

\bibitem[{Landrum(2021)}]{Landrum2021RDKit2021_03_2}
Greg Landrum. 2021.
\newblock \href {https://github.com/rdkit/rdkit/releases/tag/Release_2021_03_2} {Rdkit: Open-source cheminformatics software}.
\newblock GitHub release.

\bibitem[{Lee et~al.(2019)Lee, Keum, and Nam}]{Lee2019DeepConvDTIPO}
Ingoo Lee, Jongsoo Keum, and Hojung Nam. 2019.
\newblock {DeepConv-DTI}: Prediction of drug-target interactions via deep learning with convolution on protein sequences.
\newblock \emph{PLoS Computational Biology}, 15.

\bibitem[{Li et~al.(2024)Li, Liu, Fan, Wei, Liu, Tang, and Li}]{molregpt}
Jiatong Li, Yunqing Liu, Wenqi Fan, Xiao-Yong Wei, Hui Liu, Jiliang Tang, and Qing Li. 2024.
\newblock Empowering molecule discovery for molecule-caption translation with large language models: A chatgpt perspective.
\newblock \emph{IEEE Transactions on Knowledge and Data Engineering}.

\bibitem[{Lin(2004)}]{rouge}
Chin-Yew Lin. 2004.
\newblock Rouge: A package for automatic evaluation of summaries.
\newblock In \emph{Text summarization branches out}, pages 74--81.

\bibitem[{Lin et~al.(2022)Lin, Akin, Rao, Hie, Zhu, Lu, dos Santos~Costa, Fazel-Zarandi, Sercu, Candido et~al.}]{esm2}
Zeming Lin, Halil Akin, Roshan Rao, Brian Hie, Zhongkai Zhu, Wenting Lu, Allan dos Santos~Costa, Maryam Fazel-Zarandi, Tom Sercu, Sal Candido, et~al. 2022.
\newblock Language models of protein sequences at the scale of evolution enable accurate structure prediction.
\newblock \emph{BioRxiv}.

\bibitem[{Lipman and Pearson(1985)}]{lipman1985rapid}
David~J Lipman and William~R Pearson. 1985.
\newblock Rapid and sensitive protein similarity searches.
\newblock \emph{Science}, 227(4693):1435--1441.

\bibitem[{Liu et~al.(2015)Liu, Sun, Guan, Zheng, and Zhou}]{liu2015improving}
Hui Liu, Jianjiang Sun, Jihong Guan, Jie Zheng, and Shuigeng Zhou. 2015.
\newblock Improving compound--protein interaction prediction by building up highly credible negative samples.
\newblock \emph{Bioinformatics}, 31(12):i221--i229.

\bibitem[{Liu et~al.(2023{\natexlab{a}})Liu, Ren, and Ren}]{git-mol}
Pengfei Liu, Yiming Ren, and Zhixiang Ren. 2023{\natexlab{a}}.
\newblock \href {https://doi.org/10.48550/ARXIV.2308.06911} {Git-mol: {A} multi-modal large language model for molecular science with graph, image, and text}.
\newblock \emph{CoRR}, abs/2308.06911.

\bibitem[{Liu et~al.(2022)Liu, Wang, Liu, Lasenby, Guo, and Tang}]{graphmvp}
Shengchao Liu, Hanchen Wang, Weiyang Liu, Joan Lasenby, Hongyu Guo, and Jian Tang. 2022.
\newblock \href {https://openreview.net/forum?id=xQUe1pOKPam} {Pre-training molecular graph representation with 3d geometry}.
\newblock In \emph{The Tenth International Conference on Learning Representations, {ICLR} 2022, Virtual Event, April 25-29, 2022}. OpenReview.net.

\bibitem[{Liu et~al.(2023{\natexlab{b}})Liu, Zhu, Lu, Xu, Nie, Gitter, Xiao, Tang, Guo, and Anandkumar}]{proteindt}
Shengchao Liu, Yutao Zhu, Jiarui Lu, Zhao Xu, Weili Nie, Anthony Gitter, Chaowei Xiao, Jian Tang, Hongyu Guo, and Anima Anandkumar. 2023{\natexlab{b}}.
\newblock \href {https://doi.org/10.48550/ARXIV.2302.04611} {A text-guided protein design framework}.
\newblock \emph{CoRR}, abs/2302.04611.

\bibitem[{Liu and Low(2023)}]{goat}
Tiedong Liu and Bryan Kian~Hsiang Low. 2023.
\newblock \href {https://doi.org/10.48550/ARXIV.2305.14201} {Goat: Fine-tuned llama outperforms {GPT-4} on arithmetic tasks}.
\newblock \emph{CoRR}, abs/2305.14201.

\bibitem[{Liu et~al.(2007)Liu, Lin, Wen, Jorissen, and Gilson}]{liu2007bindingdb}
Tiqing Liu, Yuhmei Lin, Xin Wen, Robert~N Jorissen, and Michael~K Gilson. 2007.
\newblock Bindingdb: a web-accessible database of experimentally determined protein--ligand binding affinities.
\newblock \emph{Nucleic acids research}, 35(suppl\_1):D198--D201.

\bibitem[{Liu et~al.(2023{\natexlab{c}})Liu, Zhang, Xia, Wu, Xie, Qin, Zhang, and Liu}]{molxpt}
Zequn Liu, Wei Zhang, Yingce Xia, Lijun Wu, Shufang Xie, Tao Qin, Ming Zhang, and Tie{-}Yan Liu. 2023{\natexlab{c}}.
\newblock \href {https://doi.org/10.18653/v1/2023.acl-short.138} {Molxpt: Wrapping molecules with text for generative pre-training}.
\newblock In \emph{Proceedings of the 61st Annual Meeting of the Association for Computational Linguistics (Volume 2: Short Papers), {ACL} 2023, Toronto, Canada, July 9-14, 2023}, pages 1606--1616. Association for Computational Linguistics.

\bibitem[{Liu et~al.(2023{\natexlab{d}})Liu, Li, Luo, Fei, Cao, Kawaguchi, Wang, and Chua}]{molca}
Zhiyuan Liu, Sihang Li, Yanchen Luo, Hao Fei, Yixin Cao, Kenji Kawaguchi, Xiang Wang, and Tat-Seng Chua. 2023{\natexlab{d}}.
\newblock \href {https://openreview.net/forum?id=14WRhMNq7H} {Mol{CA}: Molecular graph-language modeling with cross-modal projector and uni-modal adapter}.
\newblock In \emph{The 2023 Conference on Empirical Methods in Natural Language Processing}.

\bibitem[{Loshchilov and Hutter(2019)}]{DBLP:conf/iclr/LoshchilovH19}
Ilya Loshchilov and Frank Hutter. 2019.
\newblock \href {https://openreview.net/forum?id=Bkg6RiCqY7} {Decoupled weight decay regularization}.
\newblock In \emph{7th International Conference on Learning Representations, {ICLR} 2019, New Orleans, LA, USA, May 6-9, 2019}. OpenReview.net.

\bibitem[{Luo et~al.(2023{\natexlab{a}})Luo, Huang, Hong, Yang, Zhang, Wu, and Nie}]{luo2023empowering}
Yizhen Luo, Kui Huang, Massimo Hong, Kai Yang, Jiahuan Zhang, Yushuai Wu, and Zaiqin Nie. 2023{\natexlab{a}}.
\newblock Empowering ai drug discovery with explicit and implicit knowledge.
\newblock \emph{arXiv preprint arXiv:2305.01523}.

\bibitem[{Luo et~al.(2023{\natexlab{b}})Luo, Yang, Hong, Liu, and Nie}]{molfm}
Yizhen Luo, Kai Yang, Massimo Hong, Xing~Yi Liu, and Zaiqing Nie. 2023{\natexlab{b}}.
\newblock \href {https://doi.org/10.48550/ARXIV.2307.09484} {Molfm: {A} multimodal molecular foundation model}.
\newblock \emph{CoRR}, abs/2307.09484.

\bibitem[{Luo et~al.(2023{\natexlab{c}})Luo, Zhang, Fan, Yang, Wu, Qiao, and Nie}]{biomedgpt}
Yizhen Luo, Jiahuan Zhang, Siqi Fan, Kai Yang, Yushuai Wu, Mu~Qiao, and Zaiqing Nie. 2023{\natexlab{c}}.
\newblock \href {https://doi.org/10.48550/ARXIV.2308.09442} {Biomedgpt: Open multimodal generative pre-trained transformer for biomedicine}.
\newblock \emph{CoRR}, abs/2308.09442.

\bibitem[{Miller et~al.(2009)Miller, Vandome, and McBrewster}]{miller2009levenshtein}
Frederic~P Miller, Agnes~F Vandome, and John McBrewster. 2009.
\newblock Levenshtein distance: Information theory, computer science, string (computer science), string metric, damerau? levenshtein distance, spell checker, hamming distance.

\bibitem[{Nguyen et~al.(2021)Nguyen, Le, Quinn, Nguyen, Le, and Venkatesh}]{Nguyen2020GraphDTAPD}
Thin Nguyen, Hang Le, T.~Quinn, Tri~Minh Nguyen, Thuc~Duy Le, and Svetha Venkatesh. 2021.
\newblock Graph{DTA}: Predicting drug-target binding affinity with graph neural networks.
\newblock \emph{Bioinformatics}, 37(8):1140--1147.

\bibitem[{Nogueira et~al.(2021)Nogueira, Jiang, and Lin}]{DBLP:journals/corr/abs-2102-13019}
Rodrigo~Frassetto Nogueira, Zhiying Jiang, and Jimmy Lin. 2021.
\newblock \href {http://arxiv.org/abs/2102.13019} {Investigating the limitations of the transformers with simple arithmetic tasks}.
\newblock \emph{CoRR}, abs/2102.13019.

\bibitem[{OpenAI(2023)}]{gpt4}
OpenAI. 2023.
\newblock \href {https://doi.org/10.48550/arXiv.2303.08774} {{GPT-4} technical report}.
\newblock \emph{CoRR}, abs/2303.08774.

\bibitem[{O'Shea and Nash(2015)}]{o2015introduction}
Keiron O'Shea and Ryan Nash. 2015.
\newblock An introduction to convolutional neural networks.
\newblock \emph{arXiv preprint arXiv:1511.08458}.

\bibitem[{Papineni et~al.(2002)Papineni, Roukos, Ward, and Zhu}]{bleu}
Kishore Papineni, Salim Roukos, Todd Ward, and Wei{-}Jing Zhu. 2002.
\newblock \href {https://doi.org/10.3115/1073083.1073135} {Bleu: a method for automatic evaluation of machine translation}.
\newblock In \emph{Proceedings of the 40th Annual Meeting of the Association for Computational Linguistics, July 6-12, 2002, Philadelphia, PA, {USA}}, pages 311--318. {ACL}.

\bibitem[{Pearson and Lipman(1988)}]{pearson1988improved}
William~R Pearson and David~J Lipman. 1988.
\newblock Improved tools for biological sequence comparison.
\newblock \emph{Proceedings of the National Academy of Sciences}, 85(8):2444--2448.

\bibitem[{Pei et~al.(2024)Pei, Wu, Gao, Zhu, Wang, Wang, Qin, and Yan}]{pei2024leveraging}
Qizhi Pei, Lijun Wu, Kaiyuan Gao, Jinhua Zhu, Yue Wang, Zun Wang, Tao Qin, and Rui Yan. 2024.
\newblock Leveraging biomolecule and natural language through multi-modal learning: A survey.
\newblock \emph{arXiv preprint arXiv:2403.01528}.

\bibitem[{Pei et~al.(2023)Pei, Zhang, Zhu, Wu, Gao, Wu, Xia, and Yan}]{biot5}
Qizhi Pei, Wei Zhang, Jinhua Zhu, Kehan Wu, Kaiyuan Gao, Lijun Wu, Yingce Xia, and Rui Yan. 2023.
\newblock \href {https://aclanthology.org/2023.emnlp-main.70} {{B}io{T}5: Enriching cross-modal integration in biology with chemical knowledge and natural language associations}.
\newblock In \emph{Proceedings of the 2023 Conference on Empirical Methods in Natural Language Processing}, pages 1102--1123, Singapore. Association for Computational Linguistics.

\bibitem[{Preuer et~al.(2018)Preuer, Renz, Unterthiner, Hochreiter, and Klambauer}]{DBLP:journals/jcisd/PreuerRUHK18}
Kristina Preuer, Philipp Renz, Thomas Unterthiner, Sepp Hochreiter, and G{\"{u}}nter Klambauer. 2018.
\newblock \href {https://doi.org/10.1021/acs.jcim.8b00234} {Fr{\'{e}}chet chemnet distance: {A} metric for generative models for molecules in drug discovery}.
\newblock \emph{J. Chem. Inf. Model.}, 58(9):1736--1741.

\bibitem[{Raffel et~al.(2020)Raffel, Shazeer, Roberts, Lee, Narang, Matena, Zhou, Li, and Liu}]{t5}
Colin Raffel, Noam Shazeer, Adam Roberts, Katherine Lee, Sharan Narang, Michael Matena, Yanqi Zhou, Wei Li, and Peter~J Liu. 2020.
\newblock Exploring the limits of transfer learning with a unified text-to-text transformer.
\newblock \emph{The Journal of Machine Learning Research}, 21(1):5485--5551.

\bibitem[{Rajan et~al.(2021)Rajan, Zielesny, and Steinbeck}]{stout}
Kohulan Rajan, Achim Zielesny, and Christoph Steinbeck. 2021.
\newblock \href {https://doi.org/10.1186/S13321-021-00512-4} {{STOUT:} {SMILES} to {IUPAC} names using neural machine translation}.
\newblock \emph{J. Cheminformatics}, 13(1):34.

\bibitem[{Rives et~al.(2021)Rives, Meier, Sercu, Goyal, Lin, Liu, Guo, Ott, Zitnick, Ma et~al.}]{esm}
Alexander Rives, Joshua Meier, Tom Sercu, Siddharth Goyal, Zeming Lin, Jason Liu, Demi Guo, Myle Ott, C~Lawrence Zitnick, Jerry Ma, et~al. 2021.
\newblock Biological structure and function emerge from scaling unsupervised learning to 250 million protein sequences.
\newblock \emph{Proceedings of the National Academy of Sciences}, 118(15):e2016239118.

\bibitem[{Rogers and Hahn(2010)}]{rogers2010extended}
David Rogers and Mathew Hahn. 2010.
\newblock Extended-connectivity fingerprints.
\newblock \emph{Journal of chemical information and modeling}, 50(5):742--754.

\bibitem[{Saravanan and Gautham(2015)}]{saravanan2015harnessing}
Vijayakumar Saravanan and Namasivayam Gautham. 2015.
\newblock Harnessing computational biology for exact linear b-cell epitope prediction: a novel amino acid composition-based feature descriptor.
\newblock \emph{Omics: a journal of integrative biology}, 19(10):648--658.

\bibitem[{Schneider et~al.(2015)Schneider, Sayle, and Landrum}]{DBLP:journals/jcisd/SchneiderSL15}
Nadine Schneider, Roger~A. Sayle, and Gregory~A. Landrum. 2015.
\newblock \href {https://doi.org/10.1021/acs.jcim.5b00543} {Get your atoms in order - an open-source implementation of a novel and robust molecular canonicalization algorithm}.
\newblock \emph{J. Chem. Inf. Model.}, 55(10):2111--2120.

\bibitem[{Sever et~al.(2019)Sever, Roeder, Hindle, Sussman, Black, Argentine, Manos, and Inglis}]{sever2019biorxiv}
Richard Sever, Ted Roeder, Samantha Hindle, Linda Sussman, Kevin-John Black, Janet Argentine, Wayne Manos, and John~R Inglis. 2019.
\newblock biorxiv: the preprint server for biology.
\newblock \emph{BioRxiv}, page 833400.

\bibitem[{Su et~al.(2022)Su, Du, Yang, Zhou, Li, Rao, Sun, Lu, and Wen}]{momu}
Bing Su, Dazhao Du, Zhao Yang, Yujie Zhou, Jiangmeng Li, Anyi Rao, Hao Sun, Zhiwu Lu, and Ji-Rong Wen. 2022.
\newblock A molecular multimodal foundation model associating molecule graphs with natural language.
\newblock \emph{arXiv preprint arXiv:2209.05481}.

\bibitem[{Sung et~al.(2022)Sung, Jeong, Choi, Kim, Lee, and Kang}]{sung2022bern2}
Mujeen Sung, Minbyul Jeong, Yonghwa Choi, Donghyeon Kim, Jinhyuk Lee, and Jaewoo Kang. 2022.
\newblock Bern2: an advanced neural biomedical named entity recognition and normalization tool.
\newblock \emph{Bioinformatics}, 38(20):4837--4839.

\bibitem[{Suzek et~al.(2007)Suzek, Huang, McGarvey, Mazumder, and Wu}]{suzek2007uniref}
Baris~E Suzek, Hongzhan Huang, Peter McGarvey, Raja Mazumder, and Cathy~H Wu. 2007.
\newblock Uniref: comprehensive and non-redundant uniprot reference clusters.
\newblock \emph{Bioinformatics}, 23(10):1282--1288.

\bibitem[{Taori et~al.(2023)Taori, Gulrajani, Zhang, Dubois, Li, Guestrin, Liang, and Hashimoto}]{alpaca}
Rohan Taori, Ishaan Gulrajani, Tianyi Zhang, Yann Dubois, Xuechen Li, Carlos Guestrin, Percy Liang, and Tatsunori~B. Hashimoto. 2023.
\newblock Stanford alpaca: An instruction-following llama model.
\newblock \url{https://github.com/tatsu-lab/stanford_alpaca}.

\bibitem[{Taylor et~al.(2022)Taylor, Kardas, Cucurull, Scialom, Hartshorn, Saravia, Poulton, Kerkez, and Stojnic}]{galactica}
Ross Taylor, Marcin Kardas, Guillem Cucurull, Thomas Scialom, Anthony Hartshorn, Elvis Saravia, Andrew Poulton, Viktor Kerkez, and Robert Stojnic. 2022.
\newblock \href {https://doi.org/10.48550/ARXIV.2211.09085} {Galactica: {A} large language model for science}.
\newblock \emph{CoRR}, abs/2211.09085.

\bibitem[{Touvron et~al.(2023{\natexlab{a}})Touvron, Lavril, Izacard, Martinet, Lachaux, Lacroix, Rozi{\`{e}}re, Goyal, Hambro, Azhar, Rodriguez, Joulin, Grave, and Lample}]{llama}
Hugo Touvron, Thibaut Lavril, Gautier Izacard, Xavier Martinet, Marie{-}Anne Lachaux, Timoth{\'{e}}e Lacroix, Baptiste Rozi{\`{e}}re, Naman Goyal, Eric Hambro, Faisal Azhar, Aur{\'{e}}lien Rodriguez, Armand Joulin, Edouard Grave, and Guillaume Lample. 2023{\natexlab{a}}.
\newblock \href {https://doi.org/10.48550/ARXIV.2302.13971} {Llama: Open and efficient foundation language models}.
\newblock \emph{CoRR}, abs/2302.13971.

\bibitem[{Touvron et~al.(2023{\natexlab{b}})Touvron, Martin, Stone, Albert, Almahairi, Babaei, Bashlykov, Batra, Bhargava, Bhosale, Bikel, Blecher, Canton{-}Ferrer, Chen, Cucurull, Esiobu, Fernandes, Fu, Fu, Fuller, Gao, Goswami, Goyal, Hartshorn, Hosseini, Hou, Inan, Kardas, Kerkez, Khabsa, Kloumann, Korenev, Koura, Lachaux, Lavril, Lee, Liskovich, Lu, Mao, Martinet, Mihaylov, Mishra, Molybog, Nie, Poulton, Reizenstein, Rungta, Saladi, Schelten, Silva, Smith, Subramanian, Tan, Tang, Taylor, Williams, Kuan, Xu, Yan, Zarov, Zhang, Fan, Kambadur, Narang, Rodriguez, Stojnic, Edunov, and Scialom}]{llama2}
Hugo Touvron, Louis Martin, Kevin Stone, Peter Albert, Amjad Almahairi, Yasmine Babaei, Nikolay Bashlykov, Soumya Batra, Prajjwal Bhargava, Shruti Bhosale, Dan Bikel, Lukas Blecher, Cristian Canton{-}Ferrer, Moya Chen, Guillem Cucurull, David Esiobu, Jude Fernandes, Jeremy Fu, Wenyin Fu, Brian Fuller, Cynthia Gao, Vedanuj Goswami, Naman Goyal, Anthony Hartshorn, Saghar Hosseini, Rui Hou, Hakan Inan, Marcin Kardas, Viktor Kerkez, Madian Khabsa, Isabel Kloumann, Artem Korenev, Punit~Singh Koura, Marie{-}Anne Lachaux, Thibaut Lavril, Jenya Lee, Diana Liskovich, Yinghai Lu, Yuning Mao, Xavier Martinet, Todor Mihaylov, Pushkar Mishra, Igor Molybog, Yixin Nie, Andrew Poulton, Jeremy Reizenstein, Rashi Rungta, Kalyan Saladi, Alan Schelten, Ruan Silva, Eric~Michael Smith, Ranjan Subramanian, Xiaoqing~Ellen Tan, Binh Tang, Ross Taylor, Adina Williams, Jian~Xiang Kuan, Puxin Xu, Zheng Yan, Iliyan Zarov, Yuchen Zhang, Angela Fan, Melanie Kambadur, Sharan Narang, Aur{\'{e}}lien Rodriguez, Robert Stojnic, Sergey Edunov,
  and Thomas Scialom. 2023{\natexlab{b}}.
\newblock \href {https://doi.org/10.48550/ARXIV.2307.09288} {Llama 2: Open foundation and fine-tuned chat models}.
\newblock \emph{CoRR}, abs/2307.09288.

\bibitem[{Vaswani et~al.(2017)Vaswani, Shazeer, Parmar, Uszkoreit, Jones, Gomez, Kaiser, and Polosukhin}]{vaswani2017attention}
Ashish Vaswani, Noam Shazeer, Niki Parmar, Jakob Uszkoreit, Llion Jones, Aidan~N Gomez, {\L}ukasz Kaiser, and Illia Polosukhin. 2017.
\newblock Attention is all you need.
\newblock \emph{Advances in neural information processing systems}, 30.

\bibitem[{Wang et~al.(2022)Wang, Wang, Cao, and Barati~Farimani}]{molclr}
Yuyang Wang, Jianren Wang, Zhonglin Cao, and Amir Barati~Farimani. 2022.
\newblock Molecular contrastive learning of representations via graph neural networks.
\newblock \emph{Nature Machine Intelligence}, 4(3):279--287.

\bibitem[{Weininger(1988)}]{weininger1988smiles}
David Weininger. 1988.
\newblock Smiles, a chemical language and information system. 1. introduction to methodology and encoding rules.
\newblock \emph{Journal of chemical information and computer sciences}, 28(1):31--36.

\bibitem[{Weininger et~al.(1989)Weininger, Weininger, and Weininger}]{weininger1989smiles}
David Weininger, Arthur Weininger, and Joseph~L Weininger. 1989.
\newblock Smiles. 2. algorithm for generation of unique smiles notation.
\newblock \emph{Journal of chemical information and computer sciences}, 29(2):97--101.

\bibitem[{White(2020)}]{pubmed2.0}
Jacob White. 2020.
\newblock Pubmed 2.0.
\newblock \emph{Medical reference services quarterly}, 39(4):382--387.

\bibitem[{Wu et~al.(2018)Wu, Ramsundar, Feinberg, Gomes, Geniesse, Pappu, Leswing, and Pande}]{moleculenet}
Zhenqin Wu, Bharath Ramsundar, Evan~N Feinberg, Joseph Gomes, Caleb Geniesse, Aneesh~S Pappu, Karl Leswing, and Vijay Pande. 2018.
\newblock Moleculenet: a benchmark for molecular machine learning.
\newblock \emph{Chemical science}, 9(2):513--530.

\bibitem[{Xu et~al.(2023{\natexlab{a}})Xu, Guo, Duan, and McAuley}]{baize}
Canwen Xu, Daya Guo, Nan Duan, and Julian~J. McAuley. 2023{\natexlab{a}}.
\newblock \href {https://doi.org/10.48550/ARXIV.2304.01196} {Baize: An open-source chat model with parameter-efficient tuning on self-chat data}.
\newblock \emph{CoRR}, abs/2304.01196.

\bibitem[{Xu et~al.(2023{\natexlab{b}})Xu, Woicik, Poon, Altman, and Wang}]{biotranslator}
Hanwen Xu, Addie Woicik, Hoifung Poon, Russ~B Altman, and Sheng Wang. 2023{\natexlab{b}}.
\newblock Multilingual translation for zero-shot biomedical classification using biotranslator.
\newblock \emph{Nature Communications}, 14(1):738.

\bibitem[{Xu et~al.(2022)Xu, Zhang, Lu, Zhu, Zhang, Chang, Liu, and Tang}]{peer}
Minghao Xu, Zuobai Zhang, Jiarui Lu, Zhaocheng Zhu, Yangtian Zhang, Ma~Chang, Runcheng Liu, and Jian Tang. 2022.
\newblock Peer: a comprehensive and multi-task benchmark for protein sequence understanding.
\newblock \emph{Advances in Neural Information Processing Systems}, 35:35156--35173.

\bibitem[{Yamanishi et~al.(2008)Yamanishi, Araki, Gutteridge, Honda, and Kanehisa}]{sw_score}
Yoshihiro Yamanishi, Michihiro Araki, Alex Gutteridge, Wataru Honda, and Minoru Kanehisa. 2008.
\newblock \href {https://doi.org/10.1093/BIOINFORMATICS/BTN162} {Prediction of drug-target interaction networks from the integration of chemical and genomic spaces}.
\newblock In \emph{Proceedings 16th International Conference on Intelligent Systems for Molecular Biology (ISMB), Toronto, Canada, July 19-23, 2008}, pages 232--240.

\bibitem[{You et~al.(2020)You, Chen, Sui, Chen, Wang, and Shen}]{graphcl}
Yuning You, Tianlong Chen, Yongduo Sui, Ting Chen, Zhangyang Wang, and Yang Shen. 2020.
\newblock Graph contrastive learning with augmentations.
\newblock \emph{Advances in neural information processing systems}, 33:5812--5823.

\bibitem[{Zeng et~al.(2022)Zeng, Yao, Liu, and Sun}]{kv-plm}
Zheni Zeng, Yuan Yao, Zhiyuan Liu, and Maosong Sun. 2022.
\newblock A deep-learning system bridging molecule structure and biomedical text with comprehension comparable to human professionals.
\newblock \emph{Nature communications}, 13(1):862.

\bibitem[{Zhang et~al.(2021)Zhang, Liu, Wang, Lu, and Lee}]{mgssl}
Zaixi Zhang, Qi~Liu, Hao Wang, Chengqiang Lu, and Chee-Kong Lee. 2021.
\newblock Motif-based graph self-supervised learning for molecular property prediction.
\newblock \emph{Advances in Neural Information Processing Systems}, 34:15870--15882.

\bibitem[{Zhou et~al.(2023)Zhou, Gao, Ding, Zheng, Xu, Wei, Zhang, and Ke}]{uni-mol}
Gengmo Zhou, Zhifeng Gao, Qiankun Ding, Hang Zheng, Hongteng Xu, Zhewei Wei, Linfeng Zhang, and Guolin Ke. 2023.
\newblock \href {https://openreview.net/pdf?id=6K2RM6wVqKu} {Uni-mol: {A} universal 3d molecular representation learning framework}.
\newblock In \emph{The Eleventh International Conference on Learning Representations, {ICLR} 2023, Kigali, Rwanda, May 1-5, 2023}. OpenReview.net.

\bibitem[{Zitnik et~al.(2018)Zitnik, Sosic, and Leskovec}]{zitnik2018biosnap}
Marinka Zitnik, Rok Sosic, and Jure Leskovec. 2018.
\newblock Biosnap datasets: Stanford biomedical network dataset collection.
\newblock \emph{Note: http://snap. stanford. edu/biodata Cited by}, 5(1).

\end{thebibliography}

\appendix

\begin{figure*}[t!]
    \centering
    \includegraphics[width=0.9\linewidth]{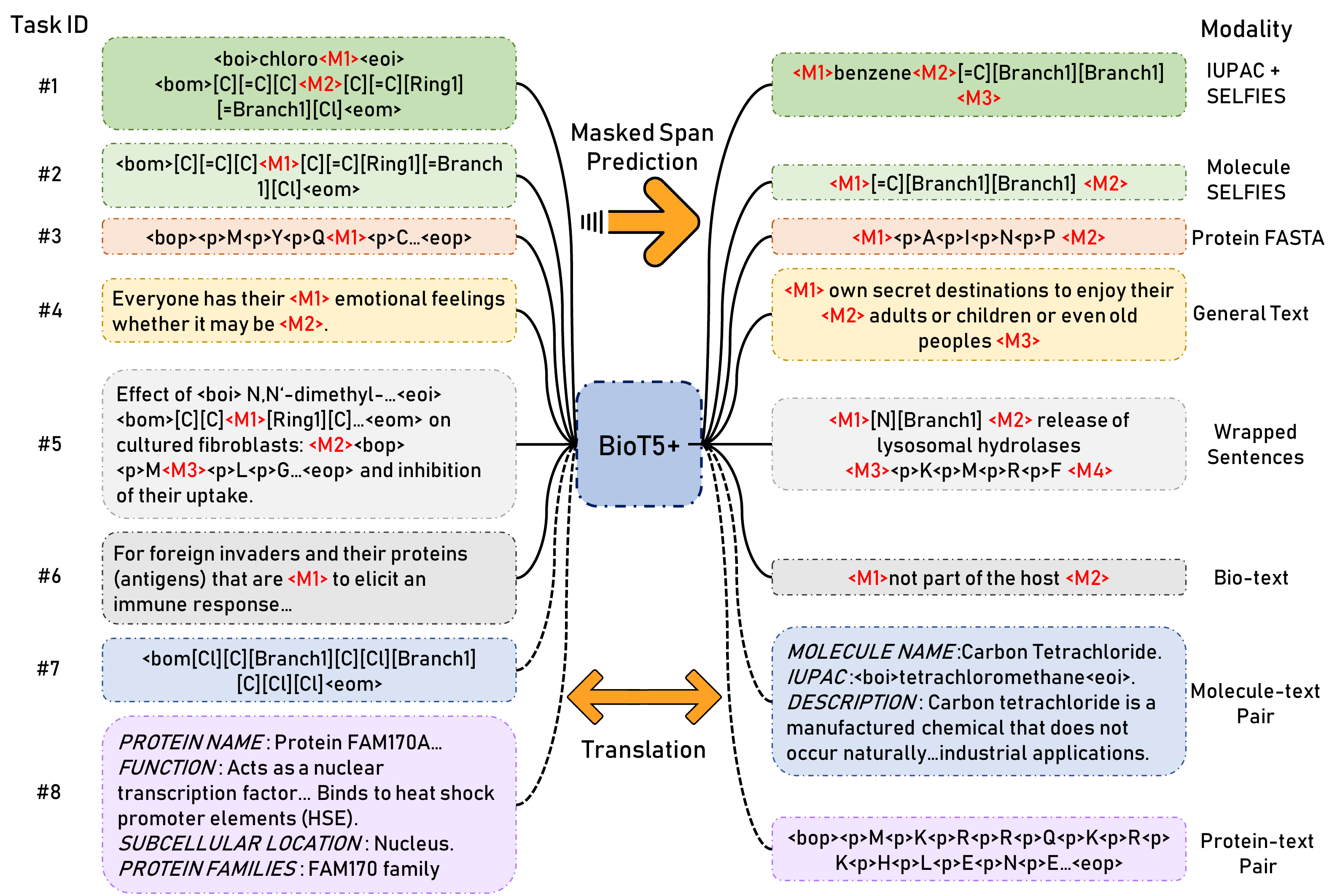}
    \caption{Overview of \method{} pre-training. The solid line refers to the masked span prediction task proposed by T5~\cite{t5}.
    Each consecutive span of masked tokens is substituted with a sentinel token, represented as \texttt{<M1>}, \texttt{<M2>}, and \texttt{<M3>}.
    We apply this pre-training task to molecule IUPAC + SELFIES (task \#1), molecule SELFIES (task \#2), protein FASTA (task \#3), general text (task \#4), wrapped text (task \#5), and bio-text (task \#6). 
    The dashed line symbolizes the bidirectional translation between structured text description and biological sequences. (task \#7 and \#8).}
    \label{fig:pipeline}
\end{figure*}

\begin{table*}[t]
\centering
\small
\caption{\footnotesize
{Performance comparison on chemical reaction-related tasks (\textbf{Best}, \underline{Second Best}). 
$*$ means LoRA tuning. 
}}
\setlength{\tabcolsep}{4.3mm}{
\scalebox{0.75}{
\begin{tabular}{lccccccc}
\toprule
\textsc{Model}
&\textsc{Exact}$\uparrow$  & \textsc{BLEU}$\uparrow$  & \textsc{Levenshtein}$\downarrow$  & \textsc{RDK FTS}$\uparrow$  & \textsc{MACCS FTS}$\uparrow$ & \textsc{Morgan FTS}$\uparrow$ & \textsc{Validity}$\uparrow$ \\
\midrule[1.1pt]
\rowcolor[RGB]{234, 238, 234}
\multicolumn{8}{l}{\textit{Reagent Prediction}} \\
Alpaca-7B & 0.000 & 0.026 & 29.037 & 0.029 & 0.016 & 0.001 & 0.186 \\
Baize-7B & 0.000 & 0.051 & 30.628 & 0.022 & 0.018 & 0.004 & 0.099 \\
ChatGLM-6B & 0.000 & 0.019 & 29.169 & 0.017 & 0.006 & 0.002 & 0.074 \\
Llama-7B & 0.000 & 0.003 & 28.040 & 0.037 & 0.001 & 0.001 & 0.001 \\
Vicuna-7B & 0.000 & 0.010 & 27.948 & 0.038 & 0.002 & 0.001 & 0.007 \\
Galactica-6.7B & 0.000 & 0.141 & 30.760 & 0.036 & 0.127 & 0.051 & 0.995 \\
Text+Chem T5-223M & 0.000 & 0.225 & 49.323 & 0.039 & 0.186 & 0.052 & 0.313 \\
Mol-Instructions-7B & 0.044 & 0.224 & 23.167 & 0.237 & 0.364 & 0.213 & 1.000 \\
Llama-7B$^*$(LoRA) &0.000	&0.283	&53.510	&0.136	&0.294	&0.106	&1.000 \\
InstructMol-G-6.9B & 0.070 & \textbf{0.890} &24.732 & \underline{0.469} & \textbf{0.691}	& \underline{0.426}	&1.000 \\
InstructMol-GS-6.9B & \underline{0.129} & 0.610 & \underline{19.664}& 0.444	&0.539	&0.400	&1.000 \\
\midrule
\textbf{\method} & \textbf{0.257} & \underline{0.695} & \textbf{12.901} & \textbf{0.539} & \underline{0.621} & \textbf{0.512} & 1.000 \\
\midrule[1.1pt]
\rowcolor[RGB]{234, 238, 234}
\multicolumn{8}{l}{\textit{Forward Reaction Prediction}} \\
Alpaca-7B & 0.000 & 0.065 & 41.989 & 0.004 & 0.024 & 0.008 & 0.138 \\
Baize-7B & 0.000 & 0.044 & 41.500 & 0.004 & 0.025 & 0.009 & 0.097 \\
ChatGLM-6B & 0.000 & 0.183 & 40.008 & 0.050 & 0.100 & 0.044 & 0.108 \\
Llama-7B & 0.000 & 0.020 & 42.002 & 0.001 & 0.002 & 0.001 & 0.039 \\
Vicuna-7B & 0.000 & 0.057 & 41.690 & 0.007 & 0.016 & 0.006 & 0.059 \\
Galactica-6.7B & 0.000 & 0.468 & 35.021 & 0.156 & 0.257 & 0.097 & 0.946 \\
Text+Chem T5-223M & 0.239 & 0.782 & 20.413 & 0.705 & 0.789 & 0.652 & 0.762 \\
Mol-Instructions-7B & 0.045 & 0.654 & 27.262 & 0.313 & 0.509 & 0.262 & 1.000 \\
Llama-7B$^*$(LoRA) &0.012	&0.804	&29.947	&0.499	&0.649	&0.407	&1.000 \\
InstructMol-G-6.9B & 0.153 &0.906 &20.155	&0.519	&0.717	&0.457	&1.000 \\
InstructMol-GS-6.9B & \underline{0.536} & \underline{0.967} & \underline{10.851} & \underline{0.776} & \underline{0.878} & \underline{0.741}	&1.000\\
\midrule
\textbf{\method} & \textbf{0.864} & \textbf{0.993} & \textbf{3.403} & \textbf{0.949} & \textbf{0.975} & \textbf{0.935} & 1.000 \\
\midrule[1.1pt]
\rowcolor[RGB]{234, 238, 234}
\multicolumn{8}{l}{\textit{Retrosynthesis}} \\
Alpaca-7B & 0.000 & 0.063 & 46.915 & 0.005 & 0.023 & 0.007 & 0.160 \\
Baize-7B & 0.000 & 0.095 & 44.714 & 0.025 & 0.050 & 0.023 & 0.112 \\
ChatGLM-6B & 0.000 & 0.117 & 48.365 & 0.056 & 0.075 & 0.043 & 0.046 \\
Llama-7B & 0.000 & 0.036 & 46.844 & 0.018 & 0.029 & 0.017 & 0.010 \\
Vicuna-7B & 0.000 & 0.057 & 46.877 & 0.025 & 0.030 & 0.021 & 0.017 \\
Galactica-6.7B & 0.000 & 0.452 & 34.940 & 0.167 & 0.274 & 0.134 & 0.986 \\
Text+Chem T5-223M & 0.141 & 0.765 & 24.043 & 0.685 & 0.765 & 0.585 & 0.698 \\
Mol-Instructions-7B & 0.009 & 0.705 & 31.227 & 0.283 & 0.487 & 0.230 & 1.000 \\
Llama-7B$^*$(LoRA) & 0.000 &0.283	&53.510	&0.136	&0.294	&0.106	&1.000 \\
InstructMol-G-6.9B & 0.114 & 0.586 & 21.271 & 0.422 & 0.523 &0.285	&1.000\\
InstructMol-GS-6.9B & \underline{0.407} & \underline{0.941} & \underline{13.967} & \underline{0.753} & \underline{0.852} & \underline{0.714} & 1.000\\
\midrule
\textbf{\method} & \textbf{0.642} & \textbf{0.969} & \textbf{6.710} & \textbf{0.897} & \textbf{0.930} & \textbf{0.866} & 1.000 \\
\bottomrule
\end{tabular}
}}
\label{tab:chemical_reaction_full}
\end{table*}

\begin{table*}[t!]
\centering
\caption{\footnotesize{
Performance comparison on the BindingDB, Human and BioSNAP datasets(\textbf{Best}, \underline{Second Best}). 
}}
\resizebox{\textwidth}{!}{
\begin{tabular}{c  ccc  cc  ccc}
\toprule
& \multicolumn{3}{c}{\textsc{BioSNAP}} & \multicolumn{2}{c}{\textsc{Human}} & \multicolumn{3}{c}{\textsc{BindingDB}} \\
\cmidrule(r){2-4}  \cmidrule(r){5-6}  \cmidrule(r){7-9} \textsc{Method} & 
\textsc{AUROC} & \textsc{AUPRC} & \textsc{Accuracy} & \textsc{AUROC} & \textsc{AUPRC} & \textsc{AUROC} & \textsc{AUPRC} & \textsc{Accuracy}\\
\midrule
\rowcolor[RGB]{234, 238, 234} \multicolumn{9}{l}{\textit{Single-task Specialist Models}} \\
SVM             & 0.862$\pm$0.007       & 0.864$\pm$0.004  & 0.777$\pm$0.011  & 0.940$\pm$0.006 & 0.920$\pm$0.009 & 0.939$\pm$0.001 & 0.928$\pm$0.002 & 0.825$\pm$0.004\\
RF        & 0.860$\pm$0.005       & 0.886$\pm$0.005  & 0.804$\pm$0.005 & 0.952$\pm$0.011 & 0.953$\pm$0.010  & 0.942$\pm$0.011        & 0.921$\pm$0.016   & 0.880$\pm$0.012\\
DeepConv-DTI           & 0.886$\pm$0.006 & 0.890$\pm$0.006  & 0.805$\pm$0.009  & 0.980$\pm$0.002 & 0.981$\pm$0.002  & 0.945$\pm$0.002       & 0.925$\pm$0.005 &0.882$\pm$0.007\\
GraphDTA      & 0.887$\pm$0.008       & 0.890$\pm$0.007  & 0.800$\pm$0.007 & 0.981$\pm$0.001 & \underline{0.982$\pm$0.002}  & 0.951$\pm$0.002        & 0.934$\pm$0.002 & 0.888$\pm$0.005\\
MolTrans & 0.895$\pm$0.004       & 0.897$\pm$0.005  & 0.825$\pm$0.010 & 0.980$\pm$0.002 & 0.978$\pm$0.003  & 0.952$\pm$0.002        & 0.936$\pm$0.001 & 0.887$\pm$0.006 \\
DrugBAN & \underline{0.903$\pm$0.005}       & \underline{0.902$\pm$0.004}  & \underline{0.834$\pm$0.008} & \underline{0.982$\pm$0.002} & 0.980$\pm$0.003 & \underline{0.960$\pm$0.001}        & \underline{0.948$\pm$0.002} & \underline{0.904$\pm$0.004}\\
BioT5 & 0.937$\pm$0.001 & 0.937$\pm$0.004 & 0.874$\pm$0.001 & \textbf{0.989$\pm$0.001} & \textbf{0.985$\pm$0.002} & 0.963$\pm$0.001 & \textbf{0.952$\pm$0.001} & \textbf{0.907$\pm$0.003}\\
\midrule
\rowcolor[RGB]{234, 238, 234} \multicolumn{9}{l}{\textit{Multi-task Generalist Models}} \\
\method & \textbf{0.939$\pm$0.001} & \textbf{0.942$\pm$0.002} & \textbf{0.875$\pm$0.001} & 0.987$\pm$0.001 & \textbf{0.985$\pm$0.002} & \textbf{0.964$\pm$0.001} & \textbf{0.952$\pm$0.001} & 0.906$\pm$0.003 \\
\bottomrule
\end{tabular}
}
\label{tab:dti_full}
\end{table*}

\section{Additional Results}
For the molecule description generation task and description-guided molecule design task, we also compare the performance of \method{} with baseline methods on Mol-Instruction~\citep{mol-instructions} test sets. The results are shown in Table~\ref{tab:molinst_mol2text} and Table~\ref{tab:molinst_text2mol}.
\method{} demonstrate superior performance in almost all metrics compared to baseline methods, which underscore \method{}'s advanced capabilities in understanding complex molecular data.

\section{Hyper-parameters}
\paragraph{Pre-training.}
The pre-training process spans $300$K steps and is executed on eight NVIDIA 80GB A100 GPUs with batch size $96$ per GPU.
To balance the data from different tasks during pre-training, we adopt a batch-level balancing strategy, where each batch evenly includes data from all eight different tasks, ensuring a more balanced and comprehensive pre-training process. For small datasets, such as molecule-text pairs and protein-text pairs, we employ a round-robin strategy to repeat their usage multiple times, compensating for their limited size.
The dropout rate is maintained at $0.0$ and the maximum input length during pre-training is set at $512$.
Optimization is performed using the AdamW~\citep{DBLP:conf/iclr/LoshchilovH19} optimizer with Root Mean Square scaling. 
A cosine annealing learning rate scheduler is employed, with the base rate set at $1e-2$ and the minimum rate at $1e-5$. 
\paragraph{Multi-task Fine-tuning.}
For multi-task fine-tuning, the dropout rate is searched in [0.0, 0.05, 0.1], and the learning rate is searched in [5e-5, 1e-4, 2e-4, 5e-4].
The total number of steps is $100$K and warmup steps is 6\% of total steps.
The batch size is set to $768$ for molecule-oriented tasks and $96$ for protein-oriented tasks.
The best hyper-parameters for molecule-oriented and protein-oriented task are shown in Table~\ref{tab:hyper_para}.

\begin{table*}[h]
    \centering
    \caption{Best hyper-parameters for multi-task instruction fine-tuning.}
    \label{tab:hyper_para}
    \scalebox{0.75}{
    \begin{tabular}{lcccc}
        \toprule
        \multirow{2}{*}{\textsc{Hyper-parameter}} & \multicolumn{2}{c}{\textsc{Molecule}}  & \multicolumn{2}{c}{\textsc{Protein}} \\
        \cmidrule(r){2-3}  \cmidrule(r){4-5} 
        & \textsc{Mol-Instructions} & \textsc{Others} & \textsc{Mol-Instructions} & \textsc{Others} \\
        \midrule
        Dropout Rate & 0.05 & 0.1 & 0.05 & 0.05 \\
        LR & 2e-4 & 5e-4 & 1e-4 & 1e-4 \\
        Batch Size & 768 & 768 & 96 & 96\\
        Steps & 100,000 & 100,000 & 100,000 & 100,000 \\
        \bottomrule
    \end{tabular}
    }
\end{table*}

\begin{table*}[t]
\caption{\footnotesize Ablation of additional data on the description-guided molecule design task.
}
\resizebox{\textwidth}{!}{
\centering
\begin{tabular}{lccccccccc}
\toprule
\textsc{Model} & \textsc{BLEU}$\uparrow$ & \textsc{Exact}$\uparrow$ & \textsc{Levenshtein}$\downarrow$ & \textsc{MACCS FTS}$\uparrow$ & \textsc{RDK FTS}$\uparrow$ & \textsc{Morgan FTS}$\uparrow$ & \textsc{FCD}$\downarrow$ & \textsc{Text2Mol}$\uparrow$ & \textsc{Validity}$\uparrow$ \\
\midrule
\method(single task) & 0.877 & 0.535 & 12.777 & 0.909 & 0.842 & 0.784 & 0.350 & 0.580 & 1.000 \\
\method(single task) wo additional data & 0.875 & 0.516 & 12.840 & 0.904 & 0.833 & 0.777 & 0.358 & 0.579 & 1.000 \\
\midrule
\method & 0.872 & 0.522 & 12.776 & 0.907 & 0.835 & 0.779 & 0.353 & 0.579 & 1.000 \\
\bottomrule
\end{tabular}
}
\label{tab:text2mol_abl}
\end{table*}
\begin{table*}[t]
\centering
\caption{Performance comparison of single-task and multi-task versions of \method{} across various datasets and tasks. \textit{Mol2text} means molecule description generation, and \textit{text2mol} means description-guided molecule design.}
\label{tab:ablation_single_multi_task}
\scalebox{0.9}{
    \begin{tabular}{lcccc}
    \toprule
    \textsc{Dataset/Task} & \textsc{Task Type} & \textsc{Single-task} & \textsc{Multi-task} & \textsc{Metric} \\
    \midrule
    BACE & classification & 87.8 & 86.2 & AUROC $\uparrow$ \\
    BBBP & classification & 78.5 & 76.5 & AUROC $\uparrow$ \\
    Clintox & classification & 97.0 & 92.3 & AUROC $\uparrow$ \\
    ChEBI-20-\textit{mol2text} & generation & 0.687 & 0.681 & METEOR $\uparrow$ \\
    ChEBI-20-\textit{text2mol} & generation & 0.877/0.535 & 0.872/0.522 & BLEU/Exact Match $\uparrow$ \\
    HOMO & regression & 0.0029 & 0.0022 & MAE $\downarrow$ \\
    LUMO & regression & 0.0029 & 0.0024 & MAE $\downarrow$ \\
    HOMO-LUMO gap & regression & 0.0040 & 0.0028 & MAE $\downarrow$ \\
    \bottomrule
    \end{tabular}
}
\end{table*}
\begin{table*}[t]
\centering
\small
\caption{\footnotesize
Performance comparison on molecule description generation task on Mol-Instrutions~\cite{mol-instructions} dataset (\textbf{Best}, \underline{Second Best}).
}
\scalebox{0.9}{
\begin{tabular}{lcccccc}
\toprule
\textsc{Model} &\textsc{BLEU-2}$\uparrow$  & \textsc{BLEU-4}$\downarrow$  & \textsc{ROUGE-1}$\uparrow$  & \textsc{ROUGE-2}$\uparrow$ & \textsc{ROUGE-L}$\uparrow$ & \textsc{METEOR}$\uparrow$ \\
\midrule 
\rowcolor[RGB]{234, 238, 234}
\multicolumn{7}{l}{\textit{Molecule Description Generation}} \\
Alpaca-7B & 0.068 & 0.014 & 0.178 & 0.041 & 0.136 & 0.107 \\
Baize-7B & 0.064 & 0.015 & 0.189 & 0.053 & 0.148 & 0.106 \\
ChatGLM-6B & 0.055 & 0.011 & 0.163 & 0.036 & 0.121 & 0.105 \\
Llama-7B & 0.059 & 0.014 & 0.164 & 0.066 & 0.148 & 0.184 \\
Vicuna-7B & 0.052 & 0.011 & 0.151 & 0.055 & 0.130 & 0.168 \\
Galactica-6.7B & 0.024 & 0.008 & 0.074 & 0.015 & 0.063 & 0.065 \\
Mol-Instructions-7B & \underline{0.217} & \underline{0.143} & \underline{0.337} & \underline{0.196} & \underline{0.291} & \underline{0.254} \\
Text+Chem T5-223M & 0.062 & 0.036 & 0.126 & 0.075 & 0.119 & 0.139 \\
MolT5-248M & 0.002 & 0.001 & 0.036 & 0.001 & 0.034 & 0.033 \\
\midrule
\textbf{\method} & \textbf{0.549} & \textbf{0.497} & \textbf{0.758} & \textbf{0.701} & \textbf{0.747} & \textbf{0.715} \\
\bottomrule
\end{tabular}
}
\label{tab:molinst_mol2text}
\end{table*}

\begin{table*}[t]
\centering
\small
\caption{\footnotesize
Performance comparison on description-guided molecule design task on Mol-Instructions~\cite{mol-instructions} dataset (\textbf{Best}, \underline{Second Best}).
}
\resizebox{\linewidth}{!}{
\begin{tabular}{lccccccc}
\toprule
\textsc{Model}
&\textsc{Exact}$\uparrow$  & \textsc{BLEU}$\uparrow$  & \textsc{Levenshtein}$\downarrow$  & \textsc{RDK FTS}$\uparrow$  & \textsc{MACCS FTS}$\uparrow$ & \textsc{Morgan FTS}$\uparrow$ & \textsc{Validity}$\uparrow$ \\
\midrule 
\rowcolor[RGB]{234, 238, 234}
\multicolumn{8}{l}{\textit{Description-guided Molecule Design}} \\
Alpaca-7B & 0.000 & 0.004 & 51.088 & 0.006 & 0.029 & 0.000 & 0.002 \\
Baize-7B & 0.000 & 0.006 & 53.796 & 0.000 & 0.000 & 0.000 & 0.002 \\
ChatGLM-6B & 0.000 & 0.004 & 53.157 & 0.005 & 0.000 & 0.000 & 0.005 \\
Llama-7B & 0.000 & 0.003 & 59.864 & 0.005 & 0.000 & 0.000 & 0.003 \\
Vicuna-7B & 0.000 & 0.006 & 60.356 & 0.006 & 0.001 & 0.000 & 0.001 \\
Galactica-6.7B & 0.000 & 0.192 & 44.152 & 0.135 & 0.248 & 0.088 & 0.992 \\
Mol-Instructions-7B &  0.002 &   0.345 &   41.367 & 0.231 &  0.412 &  0.147 &  1.000 \\
Text+Chem T5-223M & \underline{0.097} & 0.508 & 41.819 & 0.352 & 0.474 & \underline{0.353} & 0.721 \\
MolT5-248M & \textbf{0.112} & \underline{0.546} & \underline{38.276} & \underline{0.400} & \underline{0.538} & 0.295 & 0.773 \\
\midrule
\textbf{\method} & 0.079 & \textbf{0.795} & \textbf{30.728} & \textbf{0.567} & \textbf{0.687} & \textbf{0.410} & 1.000 \\
\bottomrule
\end{tabular}
}
\label{tab:molinst_text2mol}
\end{table*}

\section{NER and Entity Linking}
\label{appsec:ner_entity_link}
In general, our approach adheres to the same Named Entity Recognition (NER) and Entity Linking process as BioT5~\cite{biot5} for bio-entity name mentions in biological text using BERN2~\cite{sung2022bern2}.
However, we have implemented some modifications to the process:
(1) Upon analyzing the confidence scores of the identified bio-entities, we observed a long-tailed distribution, with the majority of bio-entity confidence scores exceeding $0.9$. Based on this empirical finding, we set a threshold of $0.9$, retaining only those NER results that surpass this score.
(2) With the introduction of IUPAC names into our workflow, we now assess whether a recognized molecular name is an IUPAC name. If it is, we exclusively append the SELFIES representation; if not, we append both the IUPAC name and the SELFIES. 
This dual approach ensures a more comprehensive and accurate representation of molecular entities in our analysis.

\section{IUPAC Incorporation}
\label{appsec:iupac}
In our downstream tasks of molecule property prediction and molecule description generation, it is effective to enrich molecules with their IUPAC name.
One reason is that IUPAC names are more commonly found in bio-text. By explicitly incorporating IUPAC names in the molecular context during pre-training, the model more readily learns relevant molecular knowledge and establishes connections between the molecule and its contextual information.
Additionally, IUPAC names inherently contain structural information about the molecule, such as functional groups and other structural components. This information allows the model to better understand the molecular structure, thus predicting molecule properties with higher accuracy and generating more accurate and detailed descriptions. 
\subsection{Mapping Process}
Initially, we normalize the SMILES sequences provided in the dataset and map them to their respective PubChem~\cite{kim2019pubchem} CIDs. Subsequently, these CIDs are used to retrieve the corresponding IUPAC names.
However, for some molecules, their SMILES sequences do not correspond to a PubChem CID. In such cases, we employ STOUT~\cite{stout}, a highly accurate SMILES to IUPAC name translator utilizing transformers, to convert these SMILES sequences into IUPAC names. 
This multi-step process ensures that each molecule is accurately equipped with its IUPAC name, facilitating more effective prediction and generation tasks.

\section{Comparison with BioT5}
Regarding the size of pre-training data and preprocessing techniques, \method{} introduces several enhancements over BioT5~\cite{biot5}:
\begin{itemize}
    \item Addition of 28.8M PubChem molecular data, including molecule SELFIES along with their IUPAC names.
    \item Addition of 28.8M full articles from PubMed.
    \item Addition of 2.3M bioRxiv abstracts as wrapped biotext.
    \item In the molecule-text bidirectional translation task, IUPAC representations are included in the text description.
    \item For wrapped biotext pre-training, detected molecule names in BioT5 are directly replaced with corresponding SELFIES. In \method, we first determine whether a molecule name is an IUPAC name; if so, it is appended with the corresponding SELFIES. If not, both the IUPAC name and SELFIES are appended. Besides, we further ensure data quality by only processing detected bioentities with a confidence score (from BERN2~\cite{sung2022bern2}) above 0.9.
    \item The numbers that appear in the pre-training corpus are tokenized character by character in BioT5+’s tokenizer.
\end{itemize}

\section{Additional Ablation Study}
\label{appsec:ablation}
To contrast single-task and multi-task tuning strategies, we further fine-tune \method{} with a single-task setting on three MoleculeNet~\cite{moleculenet} classification tasks including BACE, BBBP, Clintox, and three QM9~\cite{mol-instructions} regression tasks including HOMO, LUMO, and HOMO-LUMO gap. The consolidated results are shown in Table~\ref{tab:ablation_single_multi_task} (we also summarize the results in Table~\ref{tab:mol2text_abl} and Table~\ref{tab:text2mol_abl} for generation tasks on ChEBI-20~\cite{molt5} dataset here), covering 3 types of tasks: MoleculeNet for classification task, ChEBI-20 for generation task, and QM9 for regression task.
Our findings indicate that: (1) Single-task fine-tuning yields different performance in different tasks. For the tasks such as BACE, BBBP, Clintox, and ChEBI-20, single-task tuned BioT5+ performs closely or slightly better than the multi-task-tuned version. (2) Multi-task fine-tuning can still be advantageous for tasks with inherent correlations, such as the prediction of HOMO, LUMO, and the HOMO-LUMO gap. This evidence points to the potential for cross-task generalization, even if it is not uniformly applicable across all tasks.

\section{Fine-tuning Details}
\label{appsec:finetune_detail}
We adopt multi-task instruction tuning on molecule-oriented tasks and protein-oriented tasks.
To facilitate a fair comparison with earlier studies, given the wider variety of categories in our fine-tuning dataset compared to the Mol-Instructions~\citep{mol-instructions} dataset, we perform multi-task instruction tuning for both molecule-oriented and protein-oriented tasks, using both the Mol-Instruction dataset and an alternative dataset excluding Mol-Instructions for each sub-domain. 
Taking into account the varying sizes of the datasets involved, we report the results across differing epoch spans to accommodate these discrepancies.
All results are derived from 3 random runs.

\subsection{Molecule-oriented Tasks}
\subsubsection{Molecule Property Prediction}
\noindent{\textbf{Classification}}
We focus on the following four datasets with scaffold splits setting:

\noindent(1) BACE dataset provides both qualitative binary labels and quantitative IC50 measurements for various inhibitors aimed at human beta-secretase 1 (BACE-1).

\noindent(2) BBBP (Blood-Brain Barrier Penetration) dataset, designed to assist in predicting and modeling permeability of the blood-brain barrier, consists of compounds classified by binary labels that denote their ability to penetrate the barrier.

\noindent(3) HIV dataset includes more than 40,000 compounds evaluated for their ability to inhibit HIV replication. They were initially categorized into Confirmed Inactive (CI), Confirmed Active (CA), and Confirmed Moderately Active (CM). Later, CA and CM categories were merged, simplifying the classification into a binary system of inactive (CI) versus active (CA and CM).

\noindent(4) Clintox dataset differentiates between FDA-approved drugs and those that failed clinical trials due to toxicity. It features two distinct classification tasks with known chemical structures: (i) determining whether they exhibited toxicity in clinical trials, and (ii) assessing their FDA approval status.

We compare our \method{} with 
(1) single-task specialist models including GraphCL~\cite{graphcl}, GraphMVP-C~\cite{graphmvp}, MGSSL~\cite{mgssl}, MolCLR~\cite{molclr}, GEM~\cite{gem}, Uni-Mol~\cite{uni-mol}, KV-PLM~\cite{kv-plm}, MoMu~\cite{momu}, MolFM~\cite{molfm}, MolXPT~\cite{molxpt}, and BioT5~\cite{biot5};
(2) LLM-based generalist models including Galactica~\cite{galactica}, Vicuna~\cite{vicuna}, Llama2~\cite{llama2}, and InstructMol~\cite{instructmol}.
The baseline results are mainly derived from original papers, MolFM~\cite{molfm}, InstructMol~\cite{instructmol}.
The evaluation metric is AUROC, and we follow the same AUROC calculation method with MolXPT~\cite{molxpt} and BioT5~\cite{biot5} based on the logits of ``yes'' and ``no'' predictions.

\noindent{\textbf{Regression}}
For the regression task, we focus on three tasks from the QM9 dataset, which encompasses over 134,000 stable organic molecules with no more than nine heavy atoms, characterized by their geometric, energetic, electronic, and thermodynamic properties.
Following Mol-Instructions~\cite{mol-instructions}, our focus is on three specific subtasks within QM9:
(1) ``HOMO'' for the highest occupied molecular orbital energy.
(2) ``LUMO'' for the lowest unoccupied molecular orbital energy.
(3) ``GAP'' denoting the energy difference between HOMO and LUMO.
All of them are measured in Hartree units.
We use the processed QM9 dataset in instruction format and corresponding splits from Mol-Instruction~\cite{mol-instructions}.

We compare our \method{} with LLM-based generalist models, including Llama2~\cite{llama2}, Vicuna~\cite{vicuna}, Mol-Instructions~\cite{mol-instructions}, and InstructMol~\cite{instructmol}.
The baseline results come from InstructMol~\cite{instructmol}.
The evaluation metric is MAE.

\subsubsection{Chemical Reaction-related Tasks}
In computational chemistry, tasks centered around chemical reactions are crucial, as they significantly enhance research and development efficiency and support the advancement of eco-friendly chemistry methods. These tasks typically require understanding the interplay among reactants, reagents, and products, usually represented in the format of ``reactant > reagent > product''.
In line with~\citealp{mol-instructions} and~\citealp{instructmol}, we concentrates on three specific tasks:
(1) Reagent Prediction: This critical task entails the identification of key components such as catalysts, solvents, or auxiliary substances necessary for executing a chemical reaction. The input for this task comprises the reactants and the intended product, challenging the model to deduce the required reagents for the reaction process.
(2) Forward Reaction Prediction: This task focuses on forecasting the potential outcomes of a chemical reaction. Provided with the reactants and reagents, the objective is to accurately predict the products that would result from the chemical reaction, thereby aiding in the planning and optimization of chemical synthesis.
(3) Retrosynthesis: An essential task in synthetic chemistry, retrosynthesis involves working backwards from a target product to identify plausible reactant combinations for its synthesis. The input is a specific product compound, and the challenge lies in determining the most efficient and feasible reactants needed to produce it.
All of these chemical reaction-related data in instruction format and corresponding splits are sourced from Mol-Instruction~\cite{mol-instructions} dataset.

We compare our \method{} with LLM-based generalist models, including Alpaca~\cite{alpaca}, Baize~\cite{baize}, ChatGLM~\cite{chatglm}, Llama~\cite{llama}, Vicuna~\cite{vicuna}, Galactica~\cite{galactica}, Text+Chem T5~\cite{text+chemt5}, Mol-Instructions~\cite{mol-instructions}, and InstructMol~\cite{instructmol}.
The baseline results are sourced from InstructMol~\cite{instructmol}.
The evaluation metrics are exact match, BLEU~\cite{bleu}, Levenshtein distance~\citep{miller2009levenshtein}, three molecular fingerprints (FTS) similarity scores including MACCS~\citep{durant2002reoptimization}, RDK~\citep{DBLP:journals/jcisd/SchneiderSL15}, and Morgan~\citep{rogers2010extended}, and validity score (whether the SMILES can be successfully processed by RDKit~\citep{Landrum2021RDKit2021_03_2}).

\subsubsection{Molecule Description Generation}
\label{appsec:mol_desc_generation}
Molecule description generation task requires the model to generate a comprehensive description for a given molecule, encompassing its properties, functions, and potential applications. 
Unlike the straightforward prediction of molecular properties, this task poses a significantly greater challenge. It demands an in-depth and holistic understanding of the molecule from the model, necessitating not only the recognition of its structural and chemical characteristics but also an integration of this knowledge into a coherent, detailed narrative.
There are two well-established benchmark datasets for this task: ChEBI-20~\cite{molt5} and Mol-Instructions~\cite{mol-instructions} with recommended splits, and the corresponding results are presented in Table~\ref{tab:mol2text} and Table~\ref{tab:molinst_mol2text}.

We compare our \method{} with 
(1) single-task specialist models, including Transformer~\cite{vaswani2017attention}, T5~\cite{t5}, MolT5~\cite{molt5}, MoMu~\cite{momu}, MolFM~\cite{molfm}, MolXPT~\cite{molxpt}, GIT-Mol~\cite{git-mol}, Text+Chem T5~\cite{text+chemt5}, BioT5~\cite{biot5}, and MolCA~\cite{molca};
(2) retrieval-based LLM, including GPT-3.5-turbo~\cite{molregpt} and GPT-4~\cite{gpt4}.
(3) LLM-based generalist models, including GPT-3.5-turbo~\cite{molregpt}, BioMedGPT~\cite{biomedgpt}, Mol-Instructions~\cite{mol-instructions}, and InstructMol~\cite{instructmol}.
The baseline results are mainly derived from MolXPT~\cite{molxpt}, BioT5~\cite{biot5}, MolCA~\cite{molca}, and InstructMol~\cite{instructmol}.
The evaluation metrics are common NLP metrics, including BLEU~\cite{bleu}, ROUGE~\cite{rouge}, and METEOR~\cite{meteor}.

\subsubsection{Description-guided Molecule Design}
Description-guided molecule design is the inverse task of molecule description generation. This task centers around designing molecules based on detailed textual descriptions. Here, the model is presented with a comprehensive description encompassing various aspects of a desired molecule, such as its intended functions, properties, and potential applications. The challenge lies in accurately interpreting this textual information and translating it into a specific molecular structure. This task is considerably complex as it requires the model to have a deep understanding of the relationship between molecular characteristics and their corresponding textual descriptors.
This task is crucial for advancing drug discovery and material design, where precise molecular configurations are often derived from elaborate functional requirements.
We use the same two benchmark datasets: ChEBI-20~\cite{molt5} and Mol-Instructions~\cite{mol-instructions} as the molecule description generation task, and the corresponding results are shown in Table~\ref{tab:text2mol} and Table~\ref{tab:molinst_text2mol}.

The compared baselines are a subset of that listed in Section~\ref{appsec:mol_desc_generation} except for Llama2~\cite{llama2}, with results mainly sourced from BioT5~\cite{biot5} and MolReGPT~\cite{molregpt}.
The evaluation metrics are BLEU~\cite{bleu}, exact match, Levenshtein distance~\citep{miller2009levenshtein}, three molecular fingerprints (FTS) similarity scores including MACCS~\citep{durant2002reoptimization}, RDK~\citep{DBLP:journals/jcisd/SchneiderSL15}, and Morgan~\citep{rogers2010extended}, FCD score~\citep{DBLP:journals/jcisd/PreuerRUHK18}, Text2Mol score~\cite{text2mol}, and validity score.

\subsection{Protein-oriented Tasks}
\subsubsection{Protein Description Generation}
In computational biology, the task of protein description generation is of paramount importance, as it entails extracting insightful textual information from protein sequences. 
Aligning with~\citealp{mol-instructions}, our focus is on four intricate generation tasks, each taking a protein sequence as its input:
(1) Protein Function Generation: This task aims to produce outputs that consist of Gene Ontology (GO) terms, providing a multifaceted description of the protein's functions. These GO terms cover three key domains: cellular component, biological process, and molecular function, offering a holistic view of the protein's role and interactions within a cellular context.
(2) Catalytic Activity Generation: Here, the focus is on delineating the specific catalytic activities of the protein, moving beyond merely identifying its Enzyme Commission (EC) number. The output targets a detailed characterization of the chemical reactions facilitated by the protein, capturing its dynamic role in metabolic and biochemical pathways.
(3) Domain/Motif Generation: This task involves pinpointing and describing domains or motifs within the protein sequence. Domains and motifs are essential elements, recognized as compact, folded three-dimensional structures that play pivotal roles in the protein's function and stability. The identification of these features is crucial for understanding protein folding, function, and interactions.
(4) Functional Description Generation: The goal here is to generate a comprehensive and detailed textual description that encapsulates a protein's function, its subcellular localization, and its involvement in various biological processes. This output seeks to provide an extensive narrative, encompassing the diverse functionalities, roles, and significance of the protein within a biological system.
These tasks collectively aim to deepen our understanding of proteins, facilitating advancements in fields such as drug discovery, molecular biology, and bioinformatics.
All of these data in instruction format and corresponding splits are sourced from Mol-Instruction~\cite{mol-instructions} dataset.

We compare our \method{} with LLM-based generalist models, including Alpaca~\cite{alpaca}, Baize~\cite{baize}, ChatGLM~\cite{chatglm}, Galactica~\cite{galactica}, Llama~\cite{llama}, Vicuna~\cite{vicuna}, and Mol-Instructions~\cite{mol-instructions}.
The baseline results come from Mol-Instructions~\cite{mol-instructions}.
The evaluation metric is ROUGE-L~\cite{rouge}.

\subsection{Description-guided Protein Design}
Similar to description-guided molecule design, in the description-guided protein design task, the primary objective is to generate amino acid sequences of proteins that meet specific user-defined design requirements. 
This task necessitates a profound and comprehensive understanding of protein structures and functions from the model. 
It involves the intricate process of translating complex textual descriptions, which may include functional targets, structural characteristics, and desired biological activities, into precise amino acid sequences.
As there is no well-established benchmark for this task, we show some cases for this task in Section~\ref{appendix:text2pro_cases}.

\subsection{Protein Property Prediction}
\label{appsec:protein_prop_prediction}
The protein property prediction task plays a pivotal role in computational biology, focusing on the prediction of specific protein attributes such as solubility, structural characteristics, or functional roles, based on their amino acid sequences or structural information. 
Following~\citealp{biot5}, we centered on two key protein property prediction tasks within the PEER~\citep{peer} benchmark.
(1) Solubility Prediction: This task involves determining the solubility status of a given protein. It seeks to predict if a protein, when introduced into a solvent, will dissolve or remain insoluble. This property is crucial as it influences the protein's functionality and its interaction with other biomolecules.
(2) Localization Prediction: The second task focuses on identifying the cellular localization of proteins, distinguishing whether a given protein is ``membrane-bound'' or ``soluble''. Membrane-bound proteins are those that are associated with or integrated into the cell membrane, playing key roles in various cellular processes such as signal transduction and transport. In contrast, soluble proteins are those that are not associated with the membrane and are typically involved in various intracellular activities.
We use the same data and splitting methods as BioT5~\cite{biot5}.

We compare our \method{} with 
(1) single-task specialist models, including DDE (Dipeptide Deviation from Expected Mean)~\citep{saravanan2015harnessing}, Moran feature descriptor (Moran correlation)~\citep{feng2000prediction}, LSTM~\citep{hochreiter1997long}, Transformers~\citep{vaswani2017attention}, CNN~\citep{o2015introduction}, ResNet~\citep{he2016deep}, ProtBert~\citep{elnaggar2021prottrans}, ESM-1b~\citep{esm}, and BioT5~\cite{biot5};
(2) multi-task generalist models, including CNN~\citep{o2015introduction}, Transformers~\citep{vaswani2017attention}, and ESM-1b~\citep{esm}.
Note that here the multi-task generalist results are derived from PEER~\cite{peer}, where contact prediction, fold classification, and secondary structure prediction are combined with the original task. We report the best results obtained from training each of these tasks in conjunction with the primary task.
The baseline results are derived from PEER~\cite{peer}.
The evaluation metric is Accuracy.

\subsection{Protein-Protein Interaction}
The Protein-Protein Interaction (PPI) task is an essential component in the field of computational biology, focusing on the prediction of interactions between two proteins based on their amino acid sequences. 
In this task, the input consists of the amino acid sequences of two distinct proteins, and the output is a binary classification: ``yes'' if the proteins are predicted to interact, and ``no'' if they are not. 
This task holds substantial biological significance as protein-protein interactions are fundamental to most biological processes, including signal transduction, cellular metabolism, and immune responses. 
Following~\citealp{biot5}, we use Yeast and Human PPI datasets with corresponding splits from the PEER~\cite{peer} benchmark, which include proteins related to yeast and humans respectively. 

The compared baselines are the same with that in Section~\ref{appsec:protein_prop_prediction}, with results derived from PEER~\cite{peer}.
The evaluation metric is Accuracy.

\subsection{Drug-Target Interaction}
The Drug-Target Interaction (DTI) task focuses on the prediction of interactions between a drug molecule and a protein target. 
This task involves inputting the molecular structure of a drug, encoded as a SELFIES representation, alongside the amino acid sequence of a target protein. 
The output is a binary decision: ``yes'' indicates a predicted interaction between the drug and the protein, and ``no'' suggests no interaction. 
Understanding and predicting these interactions is crucial for drug discovery and development, offering insights into the mechanism of action of drugs, identifying potential off-target effects, and aiding in the design of novel therapeutics with improved efficacy and reduced side effects.
We use the same data and corresponding splits with BioT5~\cite{biot5}.

We compare our \method{} with single-task specialist models, including Support Vector Machine~\citep{cortes1995support} (SVM), Random Forest~\citep{ho1995random} (RF), DeepConv-DTI~\citep{Lee2019DeepConvDTIPO}, GraphDTA~\citep{Nguyen2020GraphDTAPD}, MolTrans~\citep{Huang2021MolTransMI}, DrugBAN~\citep{drugban}, and BioT5~\cite{biot5}.
The baseline results are sourced from DrugBAN~\citep{drugban} and BioT5~\cite{biot5}.
The evaluation metric is AUROC, AUPRC, and Accuracy.

\section{Case Studies}
In this section we show some cases for our fine-tuned \method{} model. 
We only showcase key parts of inputs, intentionally omitting the instruction context due to the varying requirements for instruction context across different baseline methods.
\subsection{Molecule-oriented Cases}
We show some selected cases for molecule-oriented tasks, including molecule description generation in Table~\ref{appendix:mol2text_cases}, description guided molecule design in Table~\ref{appendix:text2mol_cases}, and chemical reaction-related tasks in Table~\ref{appendix:chemical_reaction_cases}.
\begin{table*}[h!]
\centering
\small
\caption{\textbf{Molecule Description Generation Cases}. 
Given a molecule, the model needs to generate the corresponding text description.
We compare \method{} with Galactica~\cite{galactica} (6.7B), Mol-Instructions~\cite{mol-instructions}, and Text+Chem T5~\cite{text+chemt5} (base-augm version).
We adjust the instructions and molecule representations (SMILES or SELFIES) for different models according to their papers.}
\scalebox{0.92}{
\begin{tabular}{c | p{12cm} }
&  \\
\toprule
Input & [N][C][=Branch1][C][=O][N][O][C][C@H1]\\& 
[Branch1][C][NH3+1][C][=Branch1][C][=O][O-1] \\\\
Target &  
The molecule is an L-alpha-amino acid zwitterion arising from transfer of a proton from the carboxy to the amino group of O-ureido-L-serine. It is a tautomer of an O-ureido-L-serine. \\\\
\hdashline\\
Galactica &  The molecule is a zwitterion, so it has a positive charge on the nitrogen atom and a negative charge on the oxygen atom. The positive charge is on the nitrogen atom and the negative charge is on the oxygen atom. 
\\\\
Mol-Instructions & The molecule is an amino acid zwitterion obtained by transfer of a proton from the carboxy to the amino group of N-methyl-L-aspartic acid. It is an amino acid zwitterion and a N-methyl-L-aspartate. It is functionally related to a N-methyl-L-aspartic acid. It is a tautomer of a N-methyl-L-aspartic acid. It is a tautomer of a N-methyl-L-aspartic acid zw
\\\\
Text+Chem T5 & The molecule is an N-acyl-D-alpha-amino acid zwitterion obtained by transfer of a proton from the carboxy to the amino group of N-carbamoyl-D-ornithine; major species at pH 7.3. It is a tautomer of a N-carbamoyl-D-ornithine. 
\\\\
\method & The molecule is an amino acid zwitterion resulting from the transfer of a proton from the carboxy to the amino group of O-ureido-L-serine. It is a tautomer of an O-ureido-L-serine.
\\\\
\toprule
Input  &  [O][=C][Branch1][C][O][C@H1][Branch1][C][O][C][=Branch1]\\
& [C][=O][C][O][P][=Branch1][C][=O][Branch1][C][O][O] \\\\
Target &  The molecule is the ketoaldonic acid phosphate formed formally from D-erythronic acid by oxidation of the 3 -hydroxy group to an oxo group and phosphorylation at the 1 -hydroxy group. It is a conjugate acid of a (R)-2 -hydroxy-3 -oxo-4 -(phosphonatooxy)butanoate(3 -). It is an enantiomer of a (S)-2 -hydroxy-3 -oxo-4 -(phosphonooxy)butanoic acid.
\\\\
\hdashline\\
Galactica & The molecule is a carboxylic acid.
\\\\
Mol-Instructions & The molecule is a metabolite found in Escherichia coli. It has a role as a human metabolite and a Saccharomyces cerevisiae metabolite. It is a conjugate acid of a 3-phosphoglycerate(2-). It is an enantiomer of a 3-phosphoglycerate(1-). It is a tautomer of a 3-phosphoglyceric acid zwitterion. It is a tautomer of a 3-phosphoglyceric acid zwitterion.
\\\\
Text+Chem T5 & The molecule is the D-enantiomer of 2-dehydro-3-deoxy-D-glucuronic acid. It has a role as an Escherichia coli metabolite and a mouse metabolite. It is a conjugate acid of a (R)-2-dehydro-3-deoxy-D-glucuronate(2-). It is an enantiomer of a (S)-2-dehydro-3-deoxy-L-glucuronic acid.
\\\\
\method & The molecule is the ketoaldonic acid phosphate formed formally from L-erythronic acid by oxidation of the 3 -hydroxy group to an oxo group and phosphorylation at the 1 -hydroxy group. It derives from a L-erythronic acid. It is a conjugate acid of a (R)-2 -hydroxy-3 -oxo-4 -(phosphonatooxy)butanoate(3 -). It is an enantiomer of a (S)-2 -hydroxy-3 -oxo-4 -(phosphonooxy)butanoic acid.
\\\\
\toprule
Input  & [NH3+1][C@@H1][Branch1][\#C][C][C][=C][NH1][C][=C][C][=C][C]\\
& [=C][Ring1][=Branch2][Ring1][=Branch1][C][=Branch1][C][=O][O] \\\\
Target &  The molecule is the L-enantiomer of tryptophanium. It has a role as an animal metabolite and a plant metabolite. It is a conjugate acid of a L-tryptophan. It is an enantiomer of a D-tryptophanium.
\\\\
\hdashline\\
Galactica &  The molecule is [NH3+1][C@@H1](CC1=C[NH1]C2=CC=CC=C12)C(=O)O.\\\\
Mol-Instructions  & The molecule is an amino acid zwitterion. It has a role as a human metabolite. It is functionally related to an alanine. It is a tautomer of a L-alanine. It contains a L-alanine. It is a tautomer of a D-alanine zwitterion. It is an enantiomer of a D-alanine zwitterion. It is a tautomer of a D-alanine zwitterion. It is an enantiomer of a L-alanine zwitterion.
\\\\
Text+Chem T5 (Ours) & The molecule is an ammonium ion resulting from the protonation of the amino group of (S)-nefopam. It is a conjugate acid of a (S)-nefopam. It is an enantiomer of a (R)-nefopam(1+).
\\\\
\method & The molecule is the L-enantiomer of tryptophanium. It has a role as an Escherichia coli metabolite and a Saccharomyces cerevisiae metabolite. It is a conjugate acid of a L-tryptophan. It is an enantiomer of a D-tryptophanium.
\label{appendix:mol2text_cases}
\end{tabular}
}
\end{table*}

\begin{table*}[h!]
\centering
\small
\caption{\textbf{Description-guided Molecule Design}. 
Given a text description, the model needs to generate the molecule that fits the description.
We compare \method{} with Galactica~\cite{galactica} (6.7B), Mol-Instructions~\cite{mol-instructions}, and Text+Chem T5~\cite{text+chemt5} (base-augm version).
We adjust the instructions for different models according to their papers.}
\scalebox{0.99}{
\begin{tabular}{c | p{12cm} }
&  \\
\toprule
Input  & The molecule is any secondary alcohol that is one of the eight possible diastereoisomers of 5 -methyl-2 -(propan-2 -yl)cyclohexan-1 -ol. It has a role as a volatile oil component. It is a p-menthane monoterpenoid and a secondary alcohol.
\\\\
Target SMILES & CC1CCC(C(C)C)C(O)C1 \\\\
Target SELFIES & [C][C][C][C][C][Branch1][=Branch1][C][Branch1]\\
& [C][C][C][C][Branch1][C][O][C][Ring1][\#Branch2] \\\\
\hdashline \\
Galactica & CC1=CC=C(C)C(C)=C1NC(=O)C1=CC=C(Cl)C=C1
\\\\
Mol-Instructions & [C][C@@H1][C][C][C@@H1][Branch1] 
\\\\
Text+Chem T5 & CC(C)C1CCC(CC1)O
\\\\
\method{} & [C][C][C][C][C][Branch1][=Branch1][C][Branch1]\\
& [C][C][C][C][Branch1][C][O][C][Ring1][\#Branch2]
\\\\
\toprule
Input  & The molecule is a phenylsulfate oxoanion that is the conjugate base of 2 -aminophenyl hydrogen sulfate, obtained by deprotonation of the sulfo group; major species at pH 7.3. It is a conjugate base of a 2 -aminophenyl hydrogen sulfate. \\\\
Target SMILES & NC1=CC=CC=C1OS(=O)(=O)[O-1]\\\\
Target SELFIES & [N][C][=C][C][=C][C][=C][Ring1][=Branch1] \\
& [O][S][=Branch1][C][=O][=Branch1][C][=O][O-1]\\\\
\hdashline \\
Galactica & C[C@H]1C[C@@H]2[C@H]1[C@@H]1C[C@H]3[C@@H]4C[C@H](F)C5=CC\\
& (=O)C=C[C@]5(C)[C@@]4(F)[C@@H](O)C[C@]3(C)[C@]2(C(=O)CO)O1
\\\\
Mol-Instructions & [C][=C][C][=C][Branch1][Branch1][C][=C][Ring1][=Branch1]\\
& [N][S][=Branch1][C][=O][=Branch1][C][=O][O-1]
\\\\
Text+Chem T5 & C1=CC=C(C(=C1)N)OS(=O)(=O)[O-]
\\\\
\method{} & [N][C][=C][C][=C][C][=C][Ring1][=Branch1] \\
& [O][S][=Branch1][C][=O][=Branch1][C][=O][O-1]
\\\\
\toprule
Input  &  The molecule is a dicarboxylic acid. It is a conjugate acid of a diphenate(1 -). It derives from a hydride of a biphenyl.
\\\\
Target SMILES & O=C(O)C1=CC=CC=C1C2=CC=CC=C2C(=O)O\\\\
Target SELFIES & [O][=C][Branch1][C][O][C][=C][C][=C][C][=C][Ring1][=Branch1]\\
& [C][=C][C][=C][C][=C][Ring1][=Branch1][C][=Branch1][C][=O][O]\\\\
\hdashline \\
Galactica & C[C@@H]1[C@@H]2C[C@H]1[C@@H]1O[C@@H]12
\\\\
Mol-Instructions & [C][=C][C][=C][Branch1][Branch1][C][=C][Ring1][=Branch1][C][=Branch1]\\
& [C][=O][O][C][=C][C][=C][Branch1][Branch1][C][=C][Ring1][=Branch1]\\
& [C][=Branch1][C][=O][O][C][=C][C][=C][Branch1][Branch1][C][=C][Ring1][=Branch1][C]
\\\\
Text+Chem T5 & C1=CC=C(C(=C1)C2=CC=CC=C2)C(=O)O
\\\\
\method{} & [O][=C][Branch1][C][O][C][=C][C][=C][C][=C][Ring1][=Branch1]\\
& [C][=C][C][=C][C][=C][Ring1][=Branch1]
\label{appendix:text2mol_cases}
\end{tabular}
}
\end{table*}
\begin{table*}[h!]
\centering
\small
\caption{\textbf{Chemical Reaction-related Tasks}. 
We compare \method{} with Galactica~\cite{galactica} (6.7B), Mol-Instructions~\cite{mol-instructions}, and Text+Chem T5~\cite{text+chemt5} (base-augm version) on reagent prediction, forward reaction prediction, and retrosynthesis tasks.
We adjust the instructions for different models according to their papers.}
\scalebox{0.99}{
\begin{tabular}{c | p{12cm} }
&  \\
\toprule
\rowcolor[RGB]{234, 238, 234}
\textit{Reagent Prediction} \\
\hline
Input  & [O][=N+1][Branch1][C][O-1][O-1].[O][=C][NH1][C][=Branch1][C][=O][C][=C][C][=C][C][=C] \\
& [Ring1][=Branch1][NH1][Ring1][O]>>[O][=C][NH1][C][=Branch1][C][=O][C][=C][C][Branch1]\\&
[=Branch1][N+1][=Branch1][C][=O][O-1][=C][C][=C][Ring1][=Branch2][NH1][Ring1][=C]
\\\\
Target SMILES & O=S(=O)(O)O.[K+] \\\\
Target SELFIES & [O][=S][=Branch1][C][=O][Branch1][C][O][O].[K+1] \\\\
\hdashline \\
Galactica & O=C1[NH1]C(=O)C2=CC=CC=C2[NH1]1
\\\\
Mol-Instructions & [C][C][=Branch1][C][=O][O].[C][C][=Branch1][C][=O][O].[C][C][=Branch1][C][=O][O].[O]\\&[=C][Branch1][C][O-1][O-1].[O][=N+1][Branch1][C][O-1][O-1].[O][=N+1][Branch1][C][O-1][O-1].[O][=N+1][Branch1][C][O-1]
\\\\
\method{} & [O][=S][=Branch1][C][=O][Branch1][C][O][O].[K+1]
\\\\
\toprule
\rowcolor[RGB]{234, 238, 234}
\textit{Forward Reaction Prediction} \\
\hline
Input  & [C][C][=N][O][C][Branch1][=C][C][=C][C][=C][Branch1][C][Cl][C]\\&[=C][Ring1][\#Branch1][C][O][=N][Ring1][=C].[Cl][C][Cl].[O][=Mn][=O]\\\\
Target SMILES & Cc1noc(-c2ccc(Cl)cc2C=O)n1 \\\\
Target SELFIES & [C][C][=N][O][C][Branch1][=C][C][=C][C][=C][Branch1][C][Cl]\\&[C][=C][Ring1][\#Branch1][C][=O][=N][Ring1][=C]\\\\
\hdashline \\
Galactica & CC1=CC=C(C2=CC(C3=CC=C(Cl)C=C3Cl)N3NN=NC3=N2)C=C1
\\\\
Mol-Instructions & [C][C][=N][O][C][Branch1][=C][C][=C][C][=C][Branch1][C][Cl][C][=C][Ring1]
\\\\
Text+Chem T5 & CC1=NOC(C2=CC=C(Cl)C=C2C=O)=N1
\\\\
\method{} & [C][C][=N][O][C][Branch1][=C][C][=C][C][=C][Branch1][C][Cl]\\&[C][=C][Ring1][\#Branch1][C][=O][=N][Ring1][=C]
\\\\
\toprule
\rowcolor[RGB]{234, 238, 234}
\textit{Retrosynthesis} \\
\hline
Input  &  [C][C][C][C][C][=C][C][=C][Branch1][P][C][=C][C][=C][Branch1][\#Branch1][C][=Branch1]\\&[C][=O][O][C][C][=C][Ring1][\#Branch2][C][=C][Ring1][S].[C][C][C][O][C][Ring1][Branch1]\\&.[O].[Na+1].[OH1-1]
\\\\
Target SMILES & BrCc1ccccc1.Oc1ccc(I)nc1Cl\\\\
Target SELFIES & [Br][C][C][=C][C][=C][C][=C][Ring1][=Branch1].[O][C]\\&[=C][C][=C][Branch1][C][I][N][=C][Ring1][\#Branch1][Cl]\\\\
\hdashline \\
Galactica & CC1=CC=C(C2=CC=C(C(=O)O)C=C2)C=C1S(=O)(=O)[O-]
\\\\
Mol-Instructions & [C][C][C][C][C][=C][C][=C][Branch1][=Branch1][C][=C][Ring1][=Branch1]\\&[C][=C][Ring1][Branch2][C][=C][C][=C][Branch1][=Branch1][C][=Branch1]\\&[C][=O][O][C][=C][Ring1][=Branch2].[O][=C][Branch1][C][Cl][C][=Branch1][C][=O]
\\\\
Text+Chem T5 & CCCCC1=CC=C(C2=CC=C(C(=O)O)C=C2)C=C1
\\\\
\method{} & [Br][C][C][=C][C][=C][C][=C][Ring1][=Branch1].[O][C]\\&[=C][C][=C][Branch1][C][I][N][=C][Ring1][\#Branch1][Cl]
\label{appendix:chemical_reaction_cases}
\end{tabular}
}
\end{table*}

\subsection{Description-guided Protein Design}
\begin{table*}[h!]
\centering
\small
\caption{\textbf{Description-guided Protein Design}. 
Given a text description, the model needs to generate the protein that fits the description.
We compare \method{} with Galactica~\cite{galactica} (6.7B) and Mol-Instructions~\cite{mol-instructions}.
We adjust the instructions for different models according to their papers.}
\scalebox{0.99}{
\begin{tabular}{c | p{12cm} }
&  \\
\toprule
Input  & 1. The designed protein should have at least Helical transmembrane domains to ensure proper localization. \\&2. A protein that can perform RNA binding, cysteine-type endopeptidase activity, serine-type peptidase activity, ribonucleoside triphosphate phosphatase activity, structural molecule activity in viral envelope, lipid droplet, virion membrane, membrane, host cell endoplasmic reticulum membrane to support virion attachment to host cell, fusion of virus membrane with host endosome membrane, clathrin-dependent endocytosis of virus by host cell. \\&3. The protein should be designed to catalyze the following reaction a ribonucleoside 5 '-triphosphate + H2 O = a ribonucleoside 5 '- diphosphate + H(+) + phosphate. \\&4. The designed protein have Helicase ATP-binding, Peptidase C18 domains with specific sequence motifs that are crucial for its activity. \\&5. Target a Basic and acidic residues compositional bias in the protein's composition for improved properties.
\\\\
Target FASTA & MSTNPKPQRKTKRNTNRRPQDVKFPGGGQIVGGVYLLPRRGPRLGVRATRKTSERSQP\\&RGRRQPIPKARRTEGRTWAQPGYPWPLYGNEGMGWAGWLLSPRGSRPSWGPTDPRRRS\\&RNLGKVIDTLTCGFADLMGYIPLVGAPLGGAARALAHGVRVLEDGVNYATGNLPGCSFS\\&IFLLALLSCLTIPASAIEVRNVSGVYHVTNDCSNTSIVYEAADMIMRTPGCVPCVRENNSS\\&RCWVALTPTLAARNSSIP \\\\
\hdashline \\
Galactica & MRR...\\&
\textit{Normalized SW score}: 0.109
\\\\
Mol-Instructions & MSTNPKPQRKTKRNTNRRPMDVKFPGGGQIVGGVYLLPRRGPRLGVRAARKTRRSLSD\\&RESWAQLAGWRSADPTGRPEPLDWAPQRTLRGLHGAIFEPNPWPPATERALERRGHRG\\&RRSVTAGWSAREAGGTPEETPSDLSEEDDEGEDRGGTISSSPGSASPDDDAESLDTLLEP\\&AELRRRRRNLGRNGPYSGHPRYRCGCRNVDCDRPCFRCEWPCPRTVRDPCLYCRFSCTS\\&LTEEGDSGNLSNDSANSRNSVSLTAVTSAAEQQQTTNNPNLPVPIISDTILQPQMQQITS...\\&
\textit{Normalized SW score}: 0.428
\\\\
\method{} & MSTNPKPQRKTKRNTNRRPQDVKFPGGGQIVGGVYLLPRRGPRLGVRATRKTSERSQP\\&RGRRQPIPKARRPEGRTWAQPGYPWPLYGNEGCGWAGWLLSPRGSRPSWGPTDPRRRS\\&RNLGKVIDTLTCGFADLMGYIPLVGAPLGGAARALAHGVRVLEDGVNYATGNLPGCSFS\\&IFLLALLSCLTVPASAYEVRNASGVYHVTNDCSNASIVYEAADMIMHTPGCVPCVRENN\\&SSRCWVALTPTLAARNSSIP\\&
\textit{Normalized SW score}: 0.972
\\\\
\toprule
Input  &  1. The protein's functional design necessitates the inclusion of a PxxP; stabilizes the interaction of NEF/MHC-I with host AP1 M1 ; necessary for MHC-I internalization motif. \\&2. For general function, the protein need meet that Inhibits the Fas and TNFR- mediated death signals by blocking MAP3 K5 /ASK1 ; Plays a role in optimizing the host cell environment for viral replication without causing cell death by apoptosis \\&3. A protein that localizes to extracellular region, host cell plasma membrane and can perform SH3 domain binding with high specificity and efficiency.
\\\\
Target FASTA & MGGKWSKSSIVGWPAVRERIRQTEPRTEPAAEGVGAASQDLDKYGALTSSNTAHNNADC\\&AWLQAQEEEEGVGFPVKPQVPLRPMTYKAAFDLSFFLKEKGGLDGLIWSRKRQEILDLW\\&VYHTQGFFPDWQNYTPGPGVRLPLTFGWCYKLVPVDPSEVEEANKGENNCLLHPMSQY\\&GMEDEEREVLKWQFDSSLARRHLAREKHPEYYNDC\\\\
\hdashline \\
Galactica & MAA \\&
\textit{Normalized SW score}: 0.119
\\\\
Mol-Instructions & MAGHSQFKRLLTREPKVRLHGNLARSLAEQIRDWRAAEDPLYWRADPGAVEAFLRDAGE\\&PAEVVPAAAGDPTAATPEALGRGGDDDRPRSSSGGPSPRSGSGPSGAEREGGLTPDELPPPE\\&SPTSVAEEPVEEEEKEEETENSTEELSDLEEEEDSEDSGTTSLTDTTVSKKKRKSRKDKAKR\\&YKQLISKIDTEEYIKRILDCPICLDCPPDIIHPDIKDLFSDFVDLSKGVSLCDCKTDNGFTCEV\\&CGMMLIGQYSYMDPYRFIEFACSNILNRTFNSHLVPFNLMEICNDITNEIPQNFINRNTFFC...\\&
\textit{Normalized SW score}: 0.328
\\\\
\method{} & MGGKWSKSSIVGWPAVRERIRRTEPAAEGVGAASQDLDKHGALTSSNTAHNNADCAWLQ\\&AQEEEEVGFPVRPQVPLRPMTYKGAFDLSFFLKEKGGLEGLIYSKKRQEILDLWVYHTQG\\&FFPDWQNYTPGPGVRYPLTFGWCFKLVPVDPREVEEANEGENNCLLHPMSQHGMEDEDR\\&EVLKWKFDSSLARRHLAREKHPEFYKDC\\&
\textit{Normalized SW score}: 0.916
\label{appendix:text2pro_cases}
\end{tabular}
}
\end{table*}
We also show some cases of description-guided protein design task in Table~\ref{appendix:text2pro_cases}.
We compute the normalized Smith-Waterman (SW) alignment score~\citep{sw_score} between the output and ground truth protein FASTA sequence.
This method involves comparing segments of variable lengths within the protein sequences and is optimized to maximize the similarity metric, effectively evaluating the correspondence between the model output and the ground truth.
In the model output of amino acid sequences, occurrences of more than two consecutive amino acids at the terminus were manually truncated.

\end{document}